\documentclass[11pt]{article}
\usepackage[utf8]{inputenc}
\usepackage{titling}
\usepackage{authblk}
\usepackage{amssymb}
\usepackage{amsmath}
\usepackage{amsthm}
\usepackage{slashed}
\usepackage{graphicx,color,epsfig}
\usepackage{multirow}
\usepackage{subfigure}
\usepackage{cite}
\usepackage{setspace}
\usepackage[colorlinks=true,linkcolor=red,citecolor=blue]{hyperref}
\begin{document}
\title{ {New Physics Search in the Doubly Weak Decay} $\smash{\overline B}^0 \to  K^+ \pi^- $  }
\author[1,2]{Faisal Munir Bhutta}
\author[3]{Ying Li$\footnote{liying@ytu.edu.cn}$}
\author[1,2]{Cai-Dian L\"u}
\author[4]{Yue-Hong Xie}
\affil[1]{Institute of High Energy Physics, Chinese Academy of
Sciences, Beijing 100049, China}
\affil[2]{University of Chinese Academy of Sciences, Beijing 100049, China}
\affil[3]{Department of Physics, Yantai University, Yantai 264005, China}
\affil[4]{Institute of Particle Physics, Central China Normal University, Wuhan 430079, China}
\renewcommand\Authands{ and }
\maketitle
\begin{abstract}
The doubly weak  transition $b\to dd{\bar s}$  is highly suppressed in the standard model, which makes it a potential channel to explore new physics signals. We present a study of the exclusive two body wrong sign weak decay $\smash{\overline B}^0\to K^+\pi^-$ belonging to this class within the perturbative QCD approach. We perform a model independent analysis for various effective dimension-6 operators, in which large effects are possible. We further analyze the considered process in example models such as Randall-Sundrum model, including the custodially protected and the bulk-Higgs Randall-Sundrum model. Exploring the experimentally favored parameter spaces of these models leads to a large and significant enhancement of the decay rate, compared to the standard model result, which might be accessible in future experiments. We propose to look for the wrong sign decay $\smash{\overline B}^0\to K^+\pi^-$ via flavour-tagged time-dependent analyses, which can be performed at LHCb and Belle-II.
\end{abstract}
\section{Introduction}\label{Intro}
With only loop diagram contributions in the standard model (SM), rare $B$-mesons decays induced by flavor-changing neutral-current (FCNC) transitions provide an interesting possibility to explore the virtual effects from new physics (NP) beyond the SM. Various FCNC processes, which are sensitive to any new source of flavor-violating interactions, are extensively studied both within the SM and in many of the extended models. With more and more precise experimental measurements, many NP parameters are severely constrained by these channels. Example of these kinds of decays are radiative, leptonic and semi-leptonic decays, which have relatively less theoretical hadronic uncertainties.

Among the purely hadronic decays, $B\to K\pi$ decays have been studied in different NP scenarios \cite{He:1999az,Choudhury:1998wc,Liyingreview}. In the SM, the main contributions to these channels come from the penguin-induced FCNC transition $\bar b \to\bar s q \bar q$ $(q=u,d)$. Grossman {\it et al.} \cite{Grossman:1999av} have studied isospin-violating NP contributions in $B\to K\pi$ decays. Focusing on $B^\pm\to K\pi $ decays, these authors have explored the relevant observables for probing parameter spaces of different NP models. In addition, $B\to K\pi$ decays have been investigated to solve the so-called $B\to K\pi$ puzzle within the SM \cite{Li:2005kt} and in different NP models \cite{Barger:2004hn,Imbeault:2008ge}. However, from all of these hadronic rare decay studies, one can not make a definite conclusion for the new physics signals. One of the obvious reason is that the difficulty is much more than expected due to the theoretical hadronic uncertainty.

An alternative approach is the search for rare $b$ decays which have extremely small rates in the SM, so that mere detection of such processes will indicate NP. Along these lines, Huitu {\it et al.} \cite{PhysRevLett.81.4313} have suggested the processes $b\to ss\bar d$ and $b\to dd\bar s$ as prototypes. In the SM, these doubly weak transitions occur via box diagrams with up-type quarks and $W$'s inside loop, resulting in the branching ratios of approximately $\mathcal{O}(10^{-12})$ and $\mathcal{O}(10^{-14})$, respectively. Furthermore, both inclusive and exclusive channels for these transitions in different beyond SM scenarios have been investigated \cite{Huitu:1998pa, Wu:2003kp, Fajfer:2001ht, Fajfer:2000ny, Fajfer:2004fx, Fajfer:2006av}, where it is predicted that in different NP models they can be greatly enhanced. Notably, Pirjol {\it et al.} \cite{Pirjol:2009vz} have performed a systematic study of two body exclusive $B\to PP,PV, VV$ modes based on $b\to ss\bar d$ and $b\to dd\bar s$ transitions, both in the SM and for several examples of NP models such as NP with conserved global charge, minimal flavor violation (MFV), next-to-minimal flavor violation (NMFV) models and general flavor violating models.

However, measurement of these two body doubly weak decays, mediated through $b\to dd\bar s$ and $b\to ss\bar d$ transitions, is challenging, since in most cases they mix with the ordinary weak decays through $B^0_{d,s}$-$\smash{\overline B}^0_{d,s}$ mixing or $K^0$-$\smash{\overline K}^0$ mixing. In the case of $b \to dd\bar s$ transition, only suggested clear channels for experimental searches are the multi-body decays such as  $B^+\to \pi^+\pi^+K^- $ and $B_s^0\to K^-K^-\pi^+\pi^+$ decays occurring either directly or via quasi two body $PV$ $(B^+\to \pi^+\smash{\overline K}^{\ast0})$ or $VV$ $(B_s^0\to \smash{\overline K}^{\ast0}\smash{\overline K}^{\ast0})$ modes, respectively. Both $B$ factories have given the upper limit for $B^+\to \pi^+\pi^+K^-$ decay \cite{Garmash:2003er,Aubert:2008rr}, whereas the latest one is reported by the LHCb collaboration to be $\mathcal{B}(B^+\to \pi^+\pi^+K^-)<4.6\times 10^{-8}$ \cite{LHCb:2016rul}. Due to the lack of reliable QCD prediction for branching ratios of three body decays, it is difficult to interpret these upper limits as constraints to new physics parameters.
%As, theoretically, there is  no reliable QCD prediction for branching ratios of three body decays. Thus it is difficult to interpret these upper limits as constraints to new physics parameters.

In this paper, by employing the perturbative QCD factorization approach, we shall calculate the exclusive two body pure annihilation decay $\smash{\overline B}^0\to K^+\pi^-$ induced by $b\to dd\bar s$  transition. This decay can occur only through the annihilation diagrams in the SM because none of the quarks (antiquarks) in the final states are the same as those of the initial $B$ meson. It is  extremely rare in the SM. Therefore, any new physics contribution can be overwhelming. Since $B^0$ can mix with $\smash{\overline B}^0$, previously, it was thought that this channel is not distinguishable from the $B^0\to K^+\pi^-$ decay with large branching ratio at the order of $10^{-5}$. Here we would like to point out that with a large data sample one can search for the suppressed wrong sign decay by performing a flavour-tagged  time-dependent analysis, following Ref. \cite{Aubert:2004ei}. Experimentally, one can identify the $B^0$ or $\smash{\overline B}^0$ meson at the production point, e.g., by using  the charge of the lepton from the semi-leptonic decay of the other $B^0/\smash{\overline B}^0$ meson. In general, the time-dependent decay rate of an initial $\smash{\overline B}^0 (B^0)$ to the final state $K^+\pi^-$ is proportional to $\exp(-\Gamma t)[1 \mp C \cos(\Delta mt) \pm S \sin(\Delta mt)]$, where $\Delta m$ and $\Gamma$ are the mass difference and decay width of the $B^0$-$\smash{\overline B}^0$ system. Without the wrong sign decay, one expects C=1 and S=0. This can be tested at Belle-II, LHCb and its future upgrade. Any deviation from C=1 and S=0 is a signal of wrong sign $\smash{\overline B}^0\to K^+\pi^-$ decay, which may indicate a sign of new physics.
%In fact, while considering $b \to dd\bar s$ transition, this is the only two body $B\to PP$ decay channel, which can be used as a null probe of NP.

Through study, we have also found another big advantage of the wrong sign $\smash{\overline B}^0\to K^+\pi^-$ decay. The effective operators involved in the doubly weak decays, $b\to dd\bar s$ and $b\to ss\bar d$ transitions, usually also contribute to the $K^0$-$\smash{\overline K}^0$ mixing and $B^0$-$\smash{\overline B}^0$ mixing or $B_s^0$-$\smash{\overline B}_s^0$ mixing. Thus they are severely constrained by the mixing parameters  measured by the high precision experiments, which cannot contribute largely to the hadronic $B$ decays. On the contrary, in a particular example of NP such as  a model with conserved charge, some of the new physics operators contributing to the wrong sign $\smash{\overline B}^0\to K^+\pi^-$ decay through the annihilation diagram may not be severely constrained by the meson mixing parameters, due to the hierarchies among NP couplings. These operators with pseudoscalar density structure survive from helicity suppression by chiral enhancement mechanism. This can make this wrong sign decay branching ratio as large as possible to be measured by the experiments. Even in the absence of the signal, with only upper limit, the measurements of this decay will at least give the most stringent constraint to the new physics parameters.

The paper is organized as follows. In Sec.~\ref{Sec2:SM}, we calculate the decay rate for the chosen process within the SM. In Sec.~\ref{Sec3:NP}, we consider a model independent analysis of the considered channel and give predictions for the NP contributions in different NP scenarios. Next we consider that how a specific NP model with tree level FCNC transitions such as Randall-Sundrum (RS) model \cite{Randall:1999ee, Grossman:1999ra} may enhance the decay rate of the $\smash{\overline B}^0\to K^+\pi^-$ decay while satisfying all the relevant constraints. For that, we consider two models known as the RS model with custodial protection $(\text {RS}_c)$ \cite{Agashe:2006at, Carena:2006bn,Albrecht:2009xr, Biancofiore:2014wpa, Biancofiore:2014uba} in Sec.~\ref{RScmodel}, followed by the bulk-Higgs RS model \cite{Archer:2014jca} in Sec.~\ref{BulkRS}. Relevant bounds and the numerical results in the two variants of the RS model are given in Sec.~\ref{sec:06}. In Sec.~\ref{con}, we summarize our results.

%========================================================================
\section{$\smash{\overline B}^0 \to  K^+ \pi^- $ decay in the standard model}\label{Sec2:SM}
%========================================================================
The doubly weak decay $b \to dd\bar s$ transition can only occur by the box diagram in the SM that is highly suppressed. The local effective Hamiltonian for $b \to dd\bar s$ transition is give as
\begin{eqnarray}\label{01}
\mathcal{H}^{\text{SM}}=C^{\text{SM}}[(\bar d_L^{\alpha}\gamma^{\mu}b_L^{\alpha})(\bar d_L^{\beta}\gamma_{\mu}s_L^{\beta})],
\end{eqnarray}
where
\begin{eqnarray}\label{02}
C^{\text{SM}}=\frac{G_F^2m_W^2}{4\pi^2}V_{tb}V_{td}^*\Big[V_{ts}V_{td}^*f
(x)+V_{cs}V_{cd}^*\frac{m_c^2}{m_W^2}g(x,y)\Big],
\end{eqnarray}
with functions $f(x)$ and $g(x,y)$ given explicitly in \cite{PhysRevLett.81.4313}, such that $x=m_W^2/m_t^2$, $y=m_c^2/m_W^2$.

\begin{figure*}[ht]
\begin{center}
\includegraphics[width=15cm]{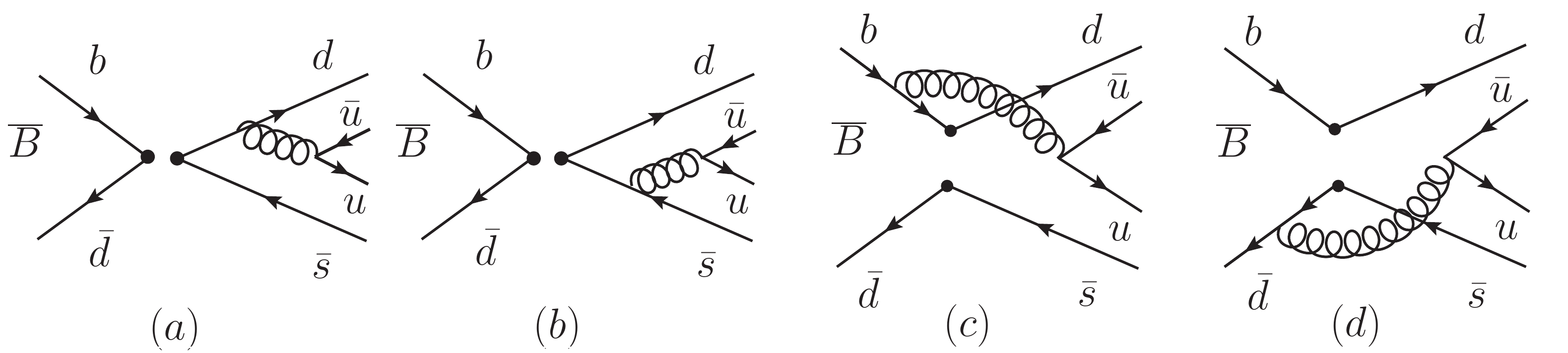}
\end{center}
\caption{Feynman diagrams which contribute to $\smash{\overline B}^0\to K^+\pi^-$ decay at leading order,  ($a$) and ($b$) are factorizable diagrams, ($c$) and ($d$) are the nonfactorizable ones.}
\label{fig:diagrams}
\end{figure*}

The exclusive $\smash{\overline B}^0\to K^+\pi^-$ decay, driven by $b \to dd\bar s$ transition, can only occur through the annihilation type Feynman diagrams. In the perturbative QCD factorization approach (PQCD), according to the effective Hamiltonian, the lowest order four annihilation Feynman diagrams for $\smash{\overline B}^0\to K^+\pi^-$ decay are shown in Fig.~\ref{fig:diagrams}, where ($a$) and ($b$) are factorizable diagrams, while ($c$) and ($d$) are the nonfactorizable ones. The initial $b$ and $\bar d$ quarks annihilate into $d$ and $\bar s$ quarks, which then form a pair of light mesons by hadronizing with another pair of $u\bar u$ produced perturbatively through the one gluon exchange mechanism.

We consider $\smash{\overline B}^0$ meson at rest for simplicity. By using the light-cone coordinates, the $\smash{\overline B}^0$ meson momentum $P_1$ and the momenta of $K^+$ and $\pi^-$ meson, denoted by $P_2$ and $P_3$, respectively can be written as
\begin{align}\label{03}
P_1=\frac{m_B}{\sqrt 2}(1,1,{\mathbf 0}_T),\qquad \;
P_2=\frac{m_B}{\sqrt 2}(1,0,{\mathbf 0}_T),\qquad \;
P_3= \frac{m_B}{\sqrt 2}(0,1,{\mathbf 0}_T).
\end{align}
The antiquark momenta $k_1$, $k_2$ and $k_3$ in the $\smash{\overline B}^0$, $K^+$ and $\pi^-$ meson  are taken as
\begin{align}\label{04}
k_1=(0,x_1P_1^{-},{\bf k}_{1T}),\qquad \;
k_2=(x_2P_2^{+},0,{\bf k}_{2T}),\qquad \;
k_3 =(0,x_3P_3^{-},{\bf k}_{3T}),
\end{align}
where the light meson masses have been neglected. In PQCD \cite{Keum:2000ph, Keum:2000wi, Lu:2000em, Wang:2006ria, Li:1992nu}, the decay amplitude is factorized into soft ($\Phi$), hard ($H$) and harder ($C$) dynamics characterized by different scales
\begin{eqnarray}\label{05}
\mathcal{A}&\sim&\int dx_1 dx_2 dx_3 b_1 db_1 b_2 db_2 b_3 db_3\nonumber
\\&&\times \mathrm{Tr}\left[C(t)\Phi_B(x_1,b_1)
\otimes\Phi_{K }(x_2,b_2)\otimes\Phi_{\pi}(x_3,b_3) H(x_i,b_i,t)S_t(x_i)e^{-S(t)}\right],
\end{eqnarray}
where $\mathrm{Tr}$ denotes the trace over Dirac and color indices, and the $b_i$ are the conjugate variables of ${\bf k}_{i T}$ of the valence quarks. The universal wave function $\Phi_M(x_i,b_i)$ ($M=B,K,\pi$), describing hadronization of the quark and anti-quark to the meson $M$, can be determined in  other decays. The explicit formulas are given in Appendix \ref{app-1}. The factorization scale $t$ is the largest energy scale in $H$, as the function in terms of $x_i$ and $b_i$. The Wilson coefficient $C(t)$ results from the radiative corrections at short distance, which includes the harder dynamics at larger scale than $m_B$ scale and describes the evolution of local 4-Fermi operators from $m_W$ down to $t$ scale.  By the threshold resummation \cite{Li:2001ay}, the large double logarithms ($\ln^2x_i$) are summed, leading to $S_t(x_i)$ which suppresses the end-point contributions. The $e^{-S(t)}$, called as Sudakov form factor \cite{Li:1997un}, contains resummation of two kinds of logarithms. One of the large logarithms is due to the renormalization of ultra-violet divergence $\ln tb$, the other one is double logarithm $\ln^2b$ from the overlap of collinear and soft gluon corrections. Such factor suppresses the soft dynamics effectively making perturbative calculation of the hard part $H$ to be reliable.

By inserting the SM operator given in Eq.~(\ref{01}), into the vertices of each Feynman diagram, we calculate the hard part $H$ at the first order of $\alpha_s$ and obtain the analytic formulas $F_{a1}$ and ${\cal M}_{a1}$, which represent the factorizable and nonfactorizable annihilation diagrams contributions, respectively. The explicit expressions of $F_{a1}$ and ${\cal M}_{a1}$ are given in the Appendix \ref{app-2}. It is obvious that these annihilation type contributions are suppressed compared to the emission diagrams, which agrees with the helicity suppression argument. Finally, the total amplitude for the considered process in the SM is given as
\begin{eqnarray}\label{06}
\mathcal{A}^{\text{SM}}= F_{a1}\bigg[\frac{4}{3}C^{\text{SM}}\bigg]+\mathcal{M}_{a1}\bigg[C^{\text{SM}}\bigg].
\end{eqnarray}
With the expression of the decay rate
\begin{eqnarray}\label{07}
\Gamma^{\text{SM}}=\frac{m_{B}^3}{64\pi}\left|\mathcal{A}^{\text{SM}}\right|^2,
\end{eqnarray}
we calculate the branching fraction of the $\smash{\overline B}^0\to K^+\pi^-$ decay in the SM
\begin{eqnarray}\label{08}
\mathcal{B}(\smash{\overline B}^0\to K^+\pi^-)^{\text{SM}}=1.0\times 10^{-19}.
\end{eqnarray}
Obviously, this order is far away from the current experimental measurement abilities, so this channel can turn out to be ideal probe of the NP effects.

%================================================================
\section{Model independent analysis of the $\smash{\overline B}^0\to K^+\pi^-$ Decay}\label{Sec3:NP}
%================================================================
In this section we present a model independent analysis of the NP contributions to the exclusive decay $\smash{\overline B}^0\to K^+\pi^-$. We start with the most general local effective Hamiltonian with all possible dimension-6 operators \cite{Grossman:1999av}
\begin{align}\label{GH} \mathcal{H}_{\text{eff}}^{\text{NP}}=\sum_{j=1}^5[C_j\mathcal{O}_j+\widetilde
{C}_j\widetilde{\mathcal{O}}_j],
\end{align}
where
\begin{gather}
\mathcal{O}_1=(\bar d_L\gamma_{\mu} b_L)(\bar d_L\gamma^{\mu}
s_L),\notag\\
\mathcal{O}_2=(\bar d_R b_L)(\bar d_R s_L),\,\,\,\,\,\,
\mathcal{O}_3=(\bar d_R^{\alpha} b_L^{\beta})(\bar d_R^{\beta}s_L^{\alpha}),\notag\\
\mathcal{O}_4=(\bar d_R b_L)(\bar d_L s_R),\,\,\,\,\,\,
\mathcal{O}_5=(\bar d_R^{\alpha} b_L^{\beta})(\bar d_L^{\beta}s_R^{\alpha}).\label{GO}
\end{gather}
The $\widetilde O_j$ operators represent the chirality flipped operators which can be obtained from $O_j$ by $L\leftrightarrow R$ exchange. In the SM only $\mathcal{O}_1$ is present. The new physics beyond SM can change the Wilson coefficient of operator $O_1$ and it can also provide non zero Wilson coefficients for other new operators. These Wilson coefficients are not free parameters, as all of them also contribute to the $K^0-\smash{\overline K}^0$ and $B^0-\smash{\overline B}^0$ mixing. The experimentally well measured mixing parameters give stringent constraints to the Wilson coefficients. This is the main reason that in many of the new physics models, one can not get too large branching ratios for the $b\to dd \bar s$ or $b\to ss \bar d$ decay, see for example \cite{Lu:2016pfs}.

We first assume that NP only gives contribution to the local operator $\mathcal{O}_1$ similar to the SM, which is true for a large class of NP models including the two-Higgs doublet model with small tan $\beta$, or the constrained minimal supersymmetry model (MSSM) \cite{Fajfer:2006av}. The NP Hamiltonian for $b \to dd\bar s$ transition in this case is the same as for the SM in Eq. (\ref{01}) but with a new Wilson coefficient $C_{1}^{dd\bar s}$.  The decay width of  $\smash{\overline B}^0\to K^+\pi^-$ decay will also have the same formula as in the SM case. The corresponding Hamiltonians for $K^0-\smash{\overline K}^0$ and $B^0-\smash{\overline B}^0$ mixing are
\begin{align}\label{12}
[\mathcal{H}_{\text{eff}}^{\Delta S=2}]=C_{1}^{K}(\bar d_L\gamma_{\mu} s_L)(\bar d_L\gamma^{\mu}
s_L),\\
[\mathcal{H}_{\text{eff}}^{\Delta B=2}]=C_{1}^{B_d}(\bar d_L\gamma_{\mu} b_L)(\bar d_L\gamma^{\mu}
b_L),
\end{align}
where the coefficients in general have the relation $C_{1}^{dd\bar s}\sim \sqrt{C_{1}^{K} C_{1}^{B_d}}$. Since the SM results of $K^0-\smash{\overline K}^0$ and $B^0-\smash{\overline B}^0$ mixing are in good agreement with the experimental data, there is not much room left for the new physics contributions.

Next, if the NP contributions come from non-standard model chiralities, considering each non-standard operator in $\mathcal{O}_{j}$ individually, the decay amplitude   is given by
\begin{eqnarray}\label{16}
\mathcal{A}_j(\smash{\overline B}^0\to K^+\pi^-)=F_{aj}\bigg[\frac{4}{3}C_j^{dd \bar s}\bigg]+\mathcal{M}_{aj}\bigg[C_j^{dd\bar s}\bigg],
\end{eqnarray}
where $j=2,3,4, 5$, respectively. The explicit expressions of $F_{a2,a3,a4,a5}$ and ${\cal M}_{a2,a3,a4,a5}$ are given in the Appendix \ref{app-2}. Similarly, for $\widetilde{\mathcal{O}}_{1-5}$, we have
\begin{eqnarray}\label{17}
\widetilde{\mathcal{A}}_j(\smash{\overline B}^0\to K^+\pi^-)=F_{aj}\bigg[\frac{4}{3}\widetilde{C}_j^{dd \bar s}\bigg]+\mathcal{M}_{aj}\bigg[-\widetilde{C}_j^{dd\bar s}\bigg],
\end{eqnarray}
The corresponding decay width of each individual operator in case of $\mathcal{O}_{2-5}$ and $\widetilde{\mathcal{O}}_{1-5}$ is given by
\begin{eqnarray}\label{18}
\Gamma_j(\smash{\overline B}^0\to K^+\pi^-)=\frac{m_{B}^3}{64\pi}\left|\mathcal{A}_j(\smash{\overline B}^0\to K^+\pi^-)\right|^2,
\end{eqnarray}
\begin{eqnarray}\label{19}
\widetilde{\Gamma}_j(\smash{\overline B}^0\to K^+\pi^-)=\frac{m_{B}^3}{64\pi}\left|\widetilde{\mathcal{A}}_j(\smash{\overline B}^0\to K^+\pi^-)\right|^2,
\end{eqnarray}
respectively. We define the ratio $R$ of the branching ratio of the wrong sign decay to the corresponding SM branching ratio of the right sign decay (induced by the $b \to su\bar u$),
\begin{eqnarray}\label{ratio}
R\equiv \frac{\mathcal{B}(\smash{\overline B}^0\to K^+\pi^-)}{\mathcal{B}(\smash{\overline B}^0\to K^-\pi^+)}.
\end{eqnarray}
We consider the corresponding LO PQCD prediction for the right sign decay, whose amplitude is given
by Eq.~(22) of \cite{Keum:2000wi} while the numerical predictions of the involved factorizable and nonfactorizable amplitudes are listed in Table.~1 of \cite{Keum:2000wi}. By direct experimental measurement of the ratio $R$, one can give constraint to each individual Wilson coefficient of the new physics operator in Eq. (\ref{GO}). For example, in Table \ref{tab:WC}, by considering each non-standard model NP operator, we obtain the upper bound on the corresponding Wilson coefficient by assuming that the current experimental precision can limit the ratio $R$ to be less than $0.001$.
\begin{table}[!tb]
 \centering \caption{Upper bounds on the Wilson coefficients of the non-standard new physics operators obtained by assuming an experimental precision of the $R<0.001$.}\label{tab:WC}
  \begin{tabular}{  c c}
    \hline\hline
    Parameter   &  Allowed range $(\text{GeV}^{-2})$   \\ \hline
       $\widetilde C_1$ & $<1.1\times 10^{-7}$\\
       $C_2$ & $<6.3\times 10^{-9}$\\
       $\widetilde C_2$ & $<6.8\times 10^{-9}$\\
       $C_3$ & $<5.1\times 10^{-8}$\\
       $\widetilde C_3$ & $<5.3\times 10^{-8}$\\
       $C_4$ & $<4.9\times 10^{-9}$\\
        $\widetilde C_4$ & $<4.2\times 10^{-9}$\\
        $C_5$ & $<1.6\times 10^{-6}$\\
         $\widetilde C_5$ & $<7.3\times 10^{-7}$ \\
    \hline\hline
  \end{tabular}
  \end{table}
%======================================================================
\subsection{NP with Conserved Charge}\label{subSec3c}
%======================================================================
We start with a NP scenario, that involves the exchange of NP fields carrying a conserved charge. A particular example of a NP scenario of this type is $R-$parity violating MSSM \cite{Fajfer:2006av}. In this class one can start with NP Lagrangian of a generic form \cite{Pirjol:2009vz}
\begin{eqnarray}\label{20}
\mathcal{L}_{\text{flavor}}=g_{b\to d}(\bar d \Gamma b)X+g_{d\to b}(\bar b \Gamma d)X+g_{s\to d}(\bar d \Gamma s)X+g_{d\to s}(\bar s \Gamma d)X+\text{h.c.},
\end{eqnarray}
assuming that field $X$ with mass $M_X$ carries a conserved quantum number broken only by the above terms. Flavor-changing operators are then obtained after integrating out the field $X$. In this example of NP, one can consider the hierarchies in NP couplings such that the mixing constraints can be
trivially satisfied, leaving the $b \to dd\bar s$ transitions unbounded. To illustrate this point, we consider four scenarios of NP. In scenario-1 (S1) and scenario-2 (S2), we consider that NP matches onto the local operators $\mathcal{O}_1$ and $\mathcal{\widetilde{O}}_1$, respectively, while in scenario-3 (S3) and scenario-4 (S4), NP matches onto the linear combination of local operators $\mathcal{O}_4$, $\mathcal{\widetilde{O}}_4$ and $\mathcal{O}_5$, $\mathcal{\widetilde{O}}_5$, respectively. As the S2, S3 and S4 involve the local operators with non-standard chiralities, so it will be convenient to define the normalized matrix elements of these non-standard operators with respect to the SM operator $\mathcal{O}_1$.
\begin{eqnarray}\label{21}
r_j(B\to M_2 M_3)\equiv \frac{\langle M_2M_3 |\mathcal{O}_j| B\rangle}{\langle M_2M_3 |\mathcal{O}_1| B\rangle},
\end{eqnarray}
where $j=2,3,4,5$. Similarly for $\mathcal{\widetilde{O}}_{1-5}$, we denote the ratio as ${\widetilde r}_j$.

Starting with S1 and S2, the NP Hamiltonians for the present case are given by $[\mathcal{H}_{\text{S1}}^{dd\bar s}]=C_1^{dd\bar s}\eta_{\text{QCD}}\mathcal{O}_1$, $[\mathcal{H}_{\text{S2}}^{dd\bar s}]={\widetilde C}_1^{dd\bar s}\eta_{\text{QCD}}\mathcal{\widetilde {O}}_1$, where $C_1^{dd\bar s}={\widetilde C}_1^{dd\bar s}=\frac{1}{M_{X}^2}\big(g_{b\to d}g_{d\to s}^{\ast}+g_{s\to d}g_{d\to b}^{\ast}\big)$. As for the RG running of the coefficients $C_1^{dd\bar s}$ and ${\widetilde C}_1^{dd\bar s}$ from the weak scale $m_W$ to $m_b$, we suppose there is no new particle within these scales and the evolution for both coefficients should be same as the case of QCD.  The evolution factor at NLO is given by $\eta_{\text{QCD}}(m_b)=0.87$. For  S3 and S4, the NP Hamiltonian and the corresponding Wilson coefficients can be written as $[\mathcal{H}_{\text{NP}}^{dd\bar s}]=f_{j\text{QCD}}\Big({C_j^{dd\bar s}\mathcal{O}_j+{\widetilde C}_j^{dd\bar s}\mathcal{\widetilde {O}}_j}\Big)$ and $C_j^{dd\bar s}=1/{M_{X}^2}(g_{b\to d}g_{d\to s}^{\ast})$, ${\widetilde C}_j^{dd\bar s}={1}/{M_{X}^2}(g_{s\to d}g_{d\to b}^{\ast})$, with $j=4, 5$ for S3 and S4, respectively. Similar to previous case, we factor out NLO QCD corrections to the Wilson coefficients in S3 and S4. However, situation is different here as operators $\mathcal{O}_4(\mathcal{\widetilde {O}}_4)$ and $\mathcal{O}_5(\mathcal{\widetilde {O}}_5)$ mix under renormalization such that the RG evolution operator is a $2\times2$ matrix, so each Wilson coefficient gets a small mixing induced contribution at NLO, which we ignore. Therefore, keeping only the dominant contributions we obtained $f_{4\text{QCD}}=2$, while $f_{5\text{QCD}}=0.9$. Considering the corresponding Hamiltonians, in all four scenarios, from $K^0-\smash{\overline K}^0$ and $B^0-\smash{\overline B}^0$ mixing we have the bounds
\begin{eqnarray}\label{boundsKB}
\frac{|g_{s\to d}g_{d\to s}^{\ast}|}{M_{X}^2}<\frac{1}{(\Lambda_j^K)^2},\qquad\;\:\frac{|g_{b\to d}g_{d\to b}^{\ast}|}{M_{X}^2}<\frac{1}{(\Lambda_j^{B_d})^2}
\end{eqnarray}
where $j=1$ for S1 and S2, while S3 and S4 correspond to $j=4, 5$, respectively. Considering the relations $|C_{j}^{v}|=1/(\Lambda_j^v)^2$, $|{\widetilde C}_{1}^{v}|=1/(\Lambda_1^v)^2$, where $v=K, B_d$, for $K^0-\smash{\overline K}^0$ and $B^0-\smash{\overline B}^0$ mixing, respectively, the lower bound on the NP scales $\Lambda_j^K$ and $\Lambda_j^{B_d}$  \cite{Bona:2007vi} are listed in Table.~\ref{tab:a}, with Im$(C_j^K)$ additionally constrained from $\epsilon_K$.
 \begin{table}[!tb]
 \centering \caption{Lower bounds on the NP scales coming from $K^0-\smash{\overline K}^0$ and $B^0-\smash{\overline B}^0$ mixing \cite{Bona:2007vi}.}\label{tab:a}
  \begin{tabular}{  c | c | c | c | c }
    \hline\hline
    \multirow{2}{*}{Operators}   &  \multicolumn{2}{c|}{$K^0-\smash{\overline K}^0$}   & \multicolumn{2}{c}{$B^0-\smash{\overline B}^0$}\\ \cline{2-5}
       & Parameter & Lower limit (TeV) & Parameter& Lower limit (TeV) \\ \hline
    $\mathcal{O}_1$ & $\Lambda_1^K$  &$1\times 10^3$ & $\Lambda_1^{B_d}$& 210 \\ \hline
    $\mathcal{O}_2$ & $\Lambda_2^K$ & $7.3\times 10^3$ & $\Lambda_2^{B_d}$&$1.2\times 10^3$  \\ \hline
    $\mathcal{O}_3$ & $\Lambda_3^K$ & $4.1\times 10^3$ & $\Lambda_3^{B_d}$&  600  \\ \hline
    $\mathcal{O}_4$ & $\Lambda_4^K$ & $17\times 10^3$ & $\Lambda_4^{B_d}$&$2.2\times 10^3$  \\ \hline
    $\mathcal{O}_5$ & $\Lambda_5^K$ & $10\times 10^3$ & $\Lambda_5^{B_d}$&$1.3\times 10^3$  \\
    \hline\hline
\end{tabular}
\end{table}

\begin{table}[!tb]
\centering \caption{Ratio of the branching fraction of the wrong sign decay to the branching fraction of the right sign decay, after satisfying mixing constraints, in the case of NP involving conserved charge $(R_X)$. $\text{S}1-\text{S}4$
represent NP scenarios corresponding to the presence of different NP operators (see text for details).}\label{tab:b}
\begin{tabular}{  c | c | c | c | c | c  }
\hline\hline
\multirow{2}{*}{Scenarios}   &  \multicolumn{4}{c|}{$R_X$} & \multirow{2}{*}{$R_{\text{SM}}$} \\ \cline{2-5}
    & $M_X$ (TeV) & Case-I  & $M_X$ (TeV) & Case-II  &\\ \hline
    S1 &\multirow{4}{*}{1.0} &0.085  & \multirow{4}{*}{10}  & $8.5\times 10^{-6}$ & $6.8\times 10^{-15}$\\ \cline{1-1}\cline{3-3}\cline{5-5}
   S2 &  & 0.074 &     &$7.3\times 10^{-6}$ & \\ \cline{1-1}\cline{3-3}\cline{5-6}
   S3 & &  55    &     & $0.005$ & \\ \cline{1-1}\cline{3-3}\cline{5-6}
   S4 & &  0.002 &     & $1.9\times 10^{-7}$ & \\
\hline\hline
\end{tabular}
\end{table}

To keep $b \to dd\bar s$ transitions unbounded and large, we take the case-I such that $g_{d\to s}=g_{b\to d}=1$, where the bounds in (\ref{boundsKB}) can be trivially satisfied, if for instance $g_{s\to d}=g_{d\to b}=0$. As mixing bounds do not constrain $M_{X}$ in this case so one can get the ratio $R_X$
as large as possible. For example, after assuming that the NP scale $M_{X}$ lies around the TeV scale, obtained ratio $R_X$ in each scenario is listed
under case-I in Table \ref{tab:b}. One can observe that the resulting ratio $R_X$, after satisfying mixing constraints, in each scenario is very large as compared to the SM result ($R_{\text{SM}}$). Therefore the measurement of ratio $R_X$, with even upper limit for the $\smash{\overline B}^0\to K^+\pi^-$ branching ratio, if no signal is found, will provide the most stringent constraints to the involved operators in this example of NP.

Further, we assume a higher NP scale at $10$ TeV. The resulting PQCD prediction for the ratio $R_X$ in corresponding scenarios is listed in Table \ref{tab:b}, as case-II. Still in scenario S3, the predicted ratio, after satisfying mixing constraints, is large enough to provide stringent constraint to the coefficient of the operator $\mathcal{O}_4$ or $\mathcal{\widetilde{O}}_{4}$. On the other hand, the obtained ratio $R_X$ in (S1, S2) and S4 give nine and eight orders of magnitude increase, respectively, compared to the SM result, although it is very difficult to reach such precision experimentally.

As a case-III, we consider a situation where accidentally one of the $g_i$ couplings is very small. For example, if we take $g_{s\to d}=0$ and all the other $g_i=1$, then $K^0-\smash{\overline K}^0$ mixing bounds in (\ref{boundsKB}) are trivially satisfied while $B^0-\smash{\overline B}^0$ mixing yields lower bound on $M_{X}$. In this case, the bound on $M_{X}$ in each scenario is more stringent compared to the previous cases, therefore the predictions for the ratio $R$  obtained for scenarios (S1, S2), S3 and S4 are of orders $10^{-11}$, $10^{-12}$ and $10^{-16}$, respectively. However compared to $R_{\text{SM}}$, as given in Table \ref{tab:b}, S1 and S2 give four and three orders of magnitude enhancement, respectively.
%==============================================
\subsection{NP Scenarios with MFV and NMFV}\label{subSec3d}
%===============================================
In MFV \cite{DAmbrosio:2002vsn}, the coefficients can be considered as $C_j^v=\frac{F_j^v}{[{(\Lambda_{\text{MFV}})}_{_j}^{v}]^2}$, where $v=dd\bar s, K, B_d$ for $b \to dd\bar s$ transitions, $K^0-\smash{\overline K}^0$ and $B^0-\smash{\overline B}^0$ mixing, respectively. In our study we restrict to the MFV case with small $\tan \beta$, where we have $F_1^v=F_{\text{SM}}^v$ and $F_{j\neq1}^v=0$. $F_{\text{SM}}^v$ are the appropriate CKM matrix elements such as, $F_{\text{SM}}^{dd\bar s}=V_{tb}V_{td}^*V_{ts}V_{td}^*$, $F_{\text{SM}}^{K}=(V_{ts}V_{td}^*)^2$ and $F_{\text{SM}}^{B_d}=(V_{tb}V_{td}^*)^2$. In this case, UTfit collaboration has given the lower bound on the MFV scale ${(\Lambda_{\text{MFV}})}_{_1}$ at 95\% probability, ${(\Lambda_{\text{MFV}})}_{_1}>5.5$ TeV \cite{Bona:2007vi}. In addition, one can define the suppression scales that include the hierarchy of the NP induced flavor changing couplings, such that we have
\begin{align}\label{25}
\frac{F_{\text{SM}}^{dd\bar s}}{[{(\Lambda_{\text{MFV}})}_{_1}^{dd\bar s}]^2}\equiv\frac{1}{(\Lambda_1^{dd\bar s})^2},\qquad \;
\frac{F_{\text{SM}}^{K}}{[{(\Lambda_{\text{MFV}})}_{_1}^{K}]^2}\equiv\frac{1}{(\Lambda_1^{K})^2},\qquad \;
\frac{F_{\text{SM}}^{B_d}}{[{(\Lambda_{\text{MFV}})}_{_1}^{B_d}]^2}\equiv\frac{1}{(\Lambda_1^{B_d})^2}.
\end{align}
With all the parameters known, one observes that the suppression scale $\Lambda_1^{dd\bar s}$ is the geometric average of the NP scales in $K^0-\smash{\overline K}^0$ and $B^0-\smash{\overline B}^0$ mixing, $\Lambda_1^{dd\bar s}= \sqrt {\Lambda_1^K \Lambda_1^{B_d}} = 458 $ TeV.

In NMFV \cite{Agashe:2005hk}, we have $|F_j^v|=F_{\text{SM}}^v$ with an arbitrary phase, the strict correlation between the Wilson coefficients is lost in this case so that we have approximately $\Lambda_j^{dd\bar s}\sim \sqrt {\Lambda_j^K \Lambda_j^{B_d}}$. Considering each operator $\mathcal{O}_j$ separately, one can obtain
the approximate value of suppression scale $\Lambda_j^{dd\bar s}$ by using the bounds from Table.~\ref{tab:a}. The corresponding PQCD predictions for the ratio $R(\mathcal{O}_j)$, with $j=1,2,3,4,5$, in case of NMFV, are of $\mathcal{O}(10^{-12})$, $\mathcal{O}(10^{-13})$, $\mathcal{O}(10^{-14})$, $\mathcal{O}(10^{-14})$ and $\mathcal{O}(10^{-17})$, respectively. Although the PQCD amplitudes for $\mathcal{\widetilde{O}}_{j}$ operators are slightly different, the predictions of the order of
magnitude for the ratios $R(\mathcal{\widetilde{O}}_{j})$ remain the same. Also, in the considered MFV case the predicted ratio is much smaller, of the order of the SM result.

The model-independent results of the UTfit group \cite{Bona:2007vi} suggested that the scale of heavy particles mediating tree-level FCNC in models of NMFV must be heavier than $\sim 60$ TeV. An application of these results to RS-type models \cite{Csaki:2008zd} showed that the measured value of $\epsilon_K$ implies that the mass scale of the lightest KK gluon must lie above $\sim 21$ TeV, if the hierarchy of the fermion masses and weak mixings is solely due to geometry and the 5D Yukawa couplings are anarchic and of $\mathcal{O}(1)$. A follow up study \cite{Blanke:2008zb}, of the $\text{RS}_c$ model, while confirming the results in \cite{Csaki:2008zd} pointed out that there exist regions in parameter space, without much fine-tuning in the 5D Yukawa couplings, which satisfy all $\Delta F=2$ and EW precision constraints for the masses of the lightest KK gauge bosons, $M_{\text{KK}}\simeq3$ TeV. Therefore, in a specific NMFV model the bounds, including the accidental cancellations among the contributions of different operators, can be weaker. In consideration to this situation, we will study $\smash{\overline B}^0\to K^+\pi^-$ decay in RS-type models below. Moreover, while considering any specific NP model one must consider the bounds additional to the ones coming from $K^0-\smash{\overline K}^0$ and $B^0-\smash{\overline B}^0$ mixing.

Finally, assuming that the scale of NP is at the mass scale probed by $K^0-\smash{\overline K}^0$ mixing, such that the corresponding $\Lambda_j^K$ are given in Table~\ref{tab:a}, and all flavor violating couplings are considered to be $\mathcal{O}(1)$. The resulting ratio $R(\mathcal{O}_1)$ is one order of magnitude larger than the SM prediction, while for all other cases, it is of the same magnitude of order as the SM value or subsequently even smaller. Similarly, for $\mathcal{\widetilde{O}}_{j}$ operators, the order of magnitude for the predictions of $R(\mathcal{\widetilde{O}}_{j})$  remains the same as that of $R(\mathcal{O}_j)$.

\section{$ \smash{\overline B}^0 \to K^+ \pi^- $ in the Custodial RS Model}\label{RScmodel}

In the RS model with custodial symmetry, we have a single warped extra dimension with the SM gauge symmetry group enlarged to the gauge group $SU(3)_c\times SU(2)_L\times SU(2)_R\times U(1)_X\times P_{LR}$ \cite{Agashe:2006at, Carena:2006bn, Albrecht:2009xr}. In the $\text{RS}_c$ model, $\smash{\overline B}^0\to K^+\pi^-$ decay, occurring through $b\to dd\bar s$ transition, is mainly effected by the tree level contributions from the lightest KK excitations of the model such as the KK gluons $\mathcal{G}^{(1)}$, KK photon $A^{(1)}$ and new heavy EW gauge bosons $(Z_H, Z^{\prime})$, while in principle $Z$ and Higgs boson should also contribute. Since $Zb_L{\bar b}_L$ coupling is protected through the discrete $P_{LR}$ symmetry in order to satisfy EW precision constraints, it causes tree-level $Z$ contributions to be negligible. Moreover, in the $\text{RS}_c$ model, $\Delta F=2$ contributions from Higgs boson exchanges are of $\mathcal O(\upsilon^4/{M_{\text{KK}}}^4)$ \cite{Bauer:2009cf}, which implies that Higgs FCNCs have limited importance. For $\Delta F=2$ observables, Higgs FCNCs provide the most prominent effects for the CP-violating parameter $\epsilon_K$, but even there compared to the KK-gluons exchange contributions they are typically smaller \cite{Duling:2009pj}. Therefore, for the considered $\text{RS}_c$ setup \cite{Albrecht:2009xr}, realizing the insignificance of the possible Higgs boson effects in $\Delta F=2$ processes, we ignore them in our study of the $\smash{\overline B}^0\to K^+\pi^-$ decay.

Further, neglecting corrections due to EW symmetry breaking and small $SU(2)_R$ breaking effects on the UV brane, we assign a common name to the masses of the first KK gauge bosons
\begin{align}\label{eq4.06}
M_{\mathcal{G}^{(1)}}=M_{Z_H}=M_{Z^{\prime}}=M_{A^{(1)}}\equiv M_{g^{(1)}}\approx2.45 M_{\text{KK}},
\end{align}
where the KK-scale, $M_{\text{KK}}\sim\mathcal{O}(\text{TeV})$, sets the mass scale for the low-lying KK excitations. The dominant contribution comes from the KK gluons $(\mathcal{G}^{(1)})$, while the new heavy EW gauge bosons $(Z_H, Z^{\prime})$ can compete with it. The KK photon $A^{(1)}$ gives very small contribution.

For $\smash{\overline B}^0\to K^+\pi^-$ decay, tree level contributions from the lightest KK gluons ${\mathcal{G}^{(1)}}$, the lightest KK photon ${A^{(1)}}$ and  $(Z_H, Z^{\prime})$ lead to the following effective Hamiltonian
\begin{align}\label{26}
[\mathcal{H}_{\text{eff}}]_{\text{RS}_c}=C_1^{VLL}\mathcal{O}_1
+C_1^{VRR}\mathcal{\widetilde{O}}_{1}+C_4^{LR}\mathcal{O}_4+C_4^{RL}\mathcal{\widetilde{O}}_{4}+C_5^{LR}\mathcal{O}_5+C_5^{RL}\mathcal{\widetilde{O}}_{5},
\end{align}
where the chosen operator basis are same as given in Eq. (\ref{GO}), and the Wilson coefficients correspond to $\mu=\mathcal{O}(M_{g^{(1)}})$. The Wilson coefficients are given by the sum
\begin{align}\label{Rswc}
C_j^{i}(M_{g^{(1)}})&=[C_j^{i}(M_{g^{(1)}})]^{\mathcal{G}^{(1)}}+[
C_j^{i}(M_{g^{(1)}})]^{A^{(1)}}+[C_j^{i}(M_{g^{(1)}})]^{Z_H,Z^{\prime}},
\end{align}
with $j=1,4,5$ and $i=VLL, VRR, LR, RL$. We point out that in the $\text{RS}_c$ model, compared to the similar processes $K^0-\smash{\overline K}^0$ and $B^0-\smash{\overline B}^0$ mixing \cite{Blanke:2008zb}, the $\smash{\overline B}^0\to K^+\pi^-$ decay receives additional contributions from the $\mathcal{\widetilde{O}}_{4}$ and $\mathcal{\widetilde{O}}_{5}$ operators. Using Fierz transformations, we calculate the contributions to the Wilson coefficients from KK gluons, denoted by $[C_j^{i}(M_{g^{(1)}})]^{\mathcal{G}^{(1)}}$ in Eq. (\ref{Rswc}), to be
\begin{align}\label{29}
[C_1^{VLL}(M_{g^{(1)}})]^{\mathcal{G}^{(1)}}&=\frac{1}{3[M_{g^{(1)}}]^2}{p_{\text{UV}}}^2[\Delta_L^{db}(\mathcal{G}^{(1)})][\Delta_L^{ds}(\mathcal{G}^{(1)})],\notag\\
[C_1^{VRR}(M_{g^{(1)}})]^{\mathcal{G}^{(1)}}&=\frac{1}{3[M_{g^{(1)}}]^2}{p_{\text{UV}}}^2[\Delta_R^{db}(\mathcal{G}^{(1)})][\Delta_R^{ds}(\mathcal{G}^{(1)})],\notag\\
[C_4^{LR}(M_{g^{(1)}})]^{\mathcal{G}^{(1)}}&=-\frac{1}{[M_{g^{(1)}}]^2}{p_{\text{UV}}}^2[\Delta_L^{db}(\mathcal{G}^{(1)})][\Delta_R^{ds}(\mathcal{G}^{(1)})],\notag\\
[C_4^{RL}(M_{g^{(1)}})]^{\mathcal{G}^{(1)}}&=-\frac{1}{[M_{g^{(1)}}]^2}{p_{\text{UV}}}^2[\Delta_R^{db}(\mathcal{G}^{(1)})][\Delta_L^{ds}(\mathcal{G}^{(1)})],\notag\\
[C_5^{LR}(M_{g^{(1)}})]^{\mathcal{G}^{(1)}}&=\frac{1}{3[M_{g^{(1)}}]^2}{p_{\text{UV}}}^2[\Delta_L^{db}(\mathcal{G}^{(1)})][\Delta_R^{ds}(\mathcal{G}^{(1)})],\notag\\
[C_5^{RL}(M_{g^{(1)}})]^{\mathcal{G}^{(1)}}&=\frac{1}{3[M_{g^{(1)}}]^2}{p_{\text{UV}}}^2[\Delta_R^{db}(\mathcal{G}^{(1)})][\Delta_L^{ds}(\mathcal{G}^{(1)})],
\end{align}
where $p_{\text{UV}}$ parameterizes the influence of brane kinetic terms on the $SU(3)_c$ coupling. We set $p_{\text{UV}}\equiv1$. Similarly, the contributions coming from KK photon $A^{(1)}$ and the new heavy EW gauge bosons $(Z_H,Z^{\prime})$, are given by
\begin{align}\label{30}
[C_1^{VLL}(M_{g^{(1)}})]^{A^{(1)}}&=\frac{1}{[M_{g^{(1)}}]^2}[\Delta_L^{db}(A^{(1)})][\Delta_L^{ds}(A^{(1)})],\notag\\
[C_1^{VRR}(M_{g^{(1)}})]^{A^{(1)}}&=\frac{1}{[M_{g^{(1)}}]^2}[\Delta_R^{db}(A^{(1)})][\Delta_R^{ds}(A^{(1)})],\notag\\
[C_5^{LR}(M_{g^{(1)}})]^{A^{(1)}}&=-\frac{2}{[M_{g^{(1)}}]^2}[\Delta_L^{db}(A^{(1)})][\Delta_R^{ds}(A^{(1)})],\notag\\
[C_5^{RL}(M_{g^{(1)}})]^{A^{(1)}}&=-\frac{2}{[M_{g^{(1)}}]^2}[\Delta_R^{db}(A^{(1)})][\Delta_L^{ds}(A^{(1)})],
\end{align}
\begin{align}\label{31}
[C_1^{VLL}(M_{g^{(1)}})]^{Z_H,Z^{\prime}}&=\frac{1}{[M_{g^{(1)}}]^2}[\Delta_L^{db}(Z^{(1)})\Delta_L^{ds}(Z^{(1)})
+\Delta_L^{db}(Z_X^{(1)})\Delta_L^{ds}(Z_X^{(1)})],\notag\\
[C_1^{VRR}(M_{g^{(1)}})]^{Z_H,Z^{\prime}}&=\frac{1}{[M_{g^{(1)}}]^2}[\Delta_R^{db}(Z^{(1)})\Delta_R^{ds}(Z^{(1)})
+\Delta_R^{db}(Z_X^{(1)})\Delta_R^{ds}(Z_X^{(1)})],\notag\\
[C_5^{LR}(M_{g^{(1)}})]^{Z_H,Z^{\prime}}&=-\frac{2}{[M_{g^{(1)}}]^2}[\Delta_L^{db}(Z^{(1)})\Delta_R^{ds}(Z^{(1)})
+\Delta_L^{db}(Z_X^{(1)})\Delta_R^{ds}(Z_X^{(1)})],\notag\\
[C_5^{RL}(M_{g^{(1)}})]^{Z_H,Z^{\prime}}&=-\frac{2}{[M_{g^{(1)}}]^2}[\Delta_R^{db}(Z^{(1)})\Delta_L^{ds}(Z^{(1)})
+\Delta_R^{db}(Z_X^{(1)})\Delta_L^{ds}(Z_X^{(1)})],
\end{align}
where the different flavor violating couplings $\Delta_{L,R}^{db}(V)$ and $\Delta_{L,R}^{ds}(V)$, with $V=\mathcal{G}^{(1)}$, $A^{(1)}$, $Z^{(1)}$, $Z_X^{(1)}$, are given in Ref. \cite{Blanke:2008zb}. These couplings involve the overlap integrals with the profiles of the zero mode fermions and shape functions of the KK gauge bosons.

Similar to Figure.~\ref{fig:diagrams}, we calculate the LO diagrams for the $\smash{\overline B}^0\to K^+\pi^-$ decay in the $\text{RS}_c$ model within the PQCD factorization approach. We adopt $(F_{a1},F_{a4},F_{a5})$ and $(\mathcal{M}_{a1},\mathcal{M}_{a4},\mathcal{M}_{a5})$ to stand for the contributions of the factorizable and nonfactorizable annihilation diagrams from the $(\mathcal{O}_1,\mathcal{\widetilde{O}}_{1}), (\mathcal{O}_4,\mathcal{\widetilde{O}}_{4})$ and $(\mathcal{O}_5,\mathcal{\widetilde{O}}_{5})$ operators, respectively. In PQCD approach Wilson coefficients are calculated at the scale $t$, which is typically of $\mathcal{O}(1-2)$ GeV. So we have employed the RG running of the WC's from the scale $M_{g^{(1)}}$ to $t$. The relevant NLO QCD factors are given in Ref. \cite{Becirevic:2001jj}. Moreover, one loop QCD and QED anomalous dimensions for the operator basis corresponding to $b\to ss\bar d$ and $b\to dd\bar s$ transitions have been derived recently \cite{Aebischer:2017gaw}. Finally, the total decay amplitude for the $\smash{\overline B}^0\to K^+\pi^-$ decay in the $\text{RS}_c$ model is given by
\begin{align}\label{32}
\mathcal{A}=&F_{a1}\left[\frac{4}{3}\left(C_1^{VLL}+C_1^{VRR}\right)\right]
+F_{a4}\left[\frac{4}{3}\left(C_4^{LR}+C_4^{RL}\right)\right]+F_{a5}\left[\frac{4}{3}\left(C_5^{LR}+C_5^{RL}\right)\right]\notag\\
&+\mathcal{M}_{a1}\Big[C_1^{VLL}-C_1^{VRR}\Big]+\mathcal{M}_{a4}\Big[C_4^{LR}-C_4^{RL}\Big]
+\mathcal{M}_{a5}\Big[C_5^{LR}-C_5^{RL}\Big],
\end{align}
where the obtained expressions of the factorization formulas $(F_{a1},F_{a4},F_{a5})$ and $(\mathcal{M}_{a1},\mathcal{M}_{a4},\mathcal{M}_{a5})$ are given in the Appendix \ref{app-2}.

%===========================================================================
\section{$\smash{\overline B}^0\to K^+\pi^-$ in the Bulk-Higgs RS model}\label{BulkRS}
%===========================================================================

The bulk-Higgs RS model is based on the 5D gauge group $SU(3)_c\times SU(2)_V\times U(1)_Y$, where all the fields are allowed to propagate in the 5D space-time including the Higgs field \cite{Archer:2014jca}. For the $\smash{\overline B}^0\to K^+\pi^-$ decay, in the bulk-Higgs RS model, we consider contributions from the tree-level exchanges of KK gluons and photons, the $Z$ boson and the Higgs boson as well as from their KK excitations and the extended scalar fields $\phi^{Z(n)}$, which are presented in the model. Further, in the bulk-Higgs RS model we consider the summation over the contributions from the whole tower of KK excitations, with the lightest KK gauge bosons states having mass $M_{g^{(1)}}\approx2.45$ $M_{\text {KK}}$. We start with the most general local effective Hamiltonian as given in Eq. (\ref{GH}), containing all possible dimension-6
operators of Eq. (\ref{GO}), and calculate the Wilson coefficients at $\mathcal{O}(M_{\text{KK}})$
\begin{align}\label{33}
C_1&=\frac{4\pi L}{M_{\text{KK}}^2}(\widetilde{\Delta}_D)_{13}\otimes(\widetilde{\Delta}_D)_{12}
\left[\frac{\alpha_s}{2}(1-\frac{1}{N_c})+\alpha Q_d^2+\frac{\alpha}{s_w^2 c_w^2}(T_3^d-Q_ds_w^2)^2\right],\notag\\
\widetilde C_1&=\frac{4\pi L}{M_{\text{KK}}^2}(\widetilde{\Delta}_d)_{13}\otimes(\widetilde{\Delta}_d)_{12}
\left[\frac{\alpha_s}{2}(1-\frac{1}{N_c})+\alpha Q_d^2+\frac{\alpha}{s_w^2 c_w^2}(-Q_d s_w^2)^2\right],\notag\\
C_4&=-\frac{4\pi L\alpha_s}{M_{\text{KK}}^2}(\widetilde{\Delta}_D)_{13}\otimes(\widetilde{\Delta}_d)_{12}-\frac{L}{\pi\beta M_{\text{KK}}^2}(\widetilde{\Omega}_d)_{13}\otimes(\widetilde{\Omega}_D)_{12},\notag\\
\widetilde C_4&=-\frac{4\pi
L\alpha_s}{M_{\text{KK}}^2}(\widetilde{\Delta}_d)_{13}\otimes(\widetilde{\Delta}_D)_{12}-\frac{L}{\pi\beta M_{\text{KK}}^2}(\widetilde{\Omega}_D)_{13}\otimes(\widetilde{\Omega}_d)_{12},\notag\\
C_5&=\frac{4\pi L}{M_{\text{KK}}^2}(\widetilde{\Delta}_D)_{13}\otimes(\widetilde{\Delta}_d)_{12}
\left[\frac{\alpha_s}{N_c}-2\alpha Q_d^2+\frac{2\alpha}{s_w^2
c_w^2}(T_3^d-Q_ds_w^2)(Q_ds_w^2)\right],\notag\\
\widetilde C_5&=\frac{4\pi L}{M_{\text{KK}}^2}(\widetilde{\Delta}_d)_{13}\otimes(\widetilde{\Delta}_D)_{12}
\left[\frac{\alpha_s}{N_c}-2\alpha Q_d^2+\frac{2\alpha}{s_w^2
c_w^2}(T_3^d-Q_ds_w^2)(Q_ds_w^2)\right],
\end{align}
where $Q_d=-1/3$, $T_3^d=-1/2$, and $N_c=3$. $\beta$ is a parameter of the model related to the Higgs profile. Higgs and scalar field $\phi^Z$ give opposite contributions to the Wilson coefficient $C_2$, thus they are cancelled by each other, giving $C_2=0$. Similarly, $\widetilde C_2=0$. The expressions of the required mixing matrices $(\widetilde{\Delta}_{D(d)})_{13}\otimes(\widetilde{\Delta}_{D(d)})_{12}$ and $(\widetilde{\Omega}_{D(d)})_{13}\otimes(\widetilde{\Omega}_{D(d)})_{12}$ in terms of the overlap integrals of boson and fermion profiles in the bulk-Higgs RS model are similar to the ones given in \cite{Bauer:2008xb,Lu:2016pfs} and are obtained by proper replacement of the involved flavors for the considered $b \to dd\bar s$ transition.

The effective Hamiltonian given in Eq. (\ref{GH}) is valid at $\mathcal{O}(M_{\text{KK}})$, which must be evolved to a low-energy scale $t$ in PQCD formalism. Hence for the evolution of the Wilson coefficients we use the formulas of the NLO QCD factors given in Ref. \cite{Becirevic:2001jj}, for the considered decay. Similar to the $\text{RS}_c$ model, LO PQCD factorization calculation of the total amplitude for the $\smash{\overline B}^0\to K^+\pi^-$ process, in the bulk Higgs RS model, yields
\begin{align}\label{34}
\mathcal{A}=&F_{a1}\left[\frac{4}{3}\left(C_1+\widetilde C_1\right)\right]
+F_{a4}\left[\frac{4}{3}\left(C_4+\widetilde C_4\right)\right]+F_{a5}\left[\frac{4}{3}\left(C_5+\widetilde C_5\right)\right]
\notag\\
&+\mathcal{M}_{a1}\Big[C_1-\widetilde C_1\Big]+\mathcal{M}_{a4}\Big[C_4-\widetilde C_4\Big]
+\mathcal{M}_{a5}\Big[C_5-\widetilde C_5\Big].
\end{align}
The involved factorization formulas are given in Appendix \ref{app-2}. Finally, the decay rate in both the RS models is given by the expression
\begin{eqnarray}\label{35}
\Gamma=\frac{m_{B}^3}{64\pi}\left|\mathcal{A}\right|^2.
\end{eqnarray}

%==============================================================
\section{Numerical Results in the RS Models}\label{sec:06}
%==============================================================
In this section we present the results of the branching ratio of the $\smash{\overline B}^0\to K^+\pi^-$ decay in both the RS models we considered. We first describe the relevant constraints on the parameter spaces of the RS models coming from the direct searches at the LHC \cite{Aad:2015fna,Sirunyan:2017uhk}, EW precision tests \cite{Malm:2013jia, Archer:2014jca, Dillon:2014zea, Malm:2014gha}, the measurements of the Higgs signal strengths at the LHC \cite{Archer:2014jca,Malm:2014gha} and from $\Delta F=2$ flavor observables \cite{Blanke:2008zb}.

In the direct search for the KK gluons resonances, through decay into $t\bar t$ pair, recent measurements at the LHC have constrained the lightest KK gluon mass $M_{g^{(1)}}>3.3$ TeV at $95\%$ confidence level \cite{Sirunyan:2017uhk}. Further, in the $\text{RS}_c$ model, EW precision constraints imposed by the tree-level analysis of the $S$ and $T$ parameters lead to $M_{g^{(1)}}>4.8$ TeV for the lowest KK gluons and KK photon masses \cite{Malm:2013jia}. Similarly, in the bulk-Higgs RS model, KK scale $(M_{\text{KK}})$ is constrained by the analyses of the EW precision data \cite{Archer:2014jca}, such that under a constrained fit (i.e. $U=0$), the obtained lower bounds on the KK mass scale vary between $M_{\text{KK}}>3.0$ TeV for $\beta=0$ to $M_{\text{KK}}>5.1$ TeV for $\beta=10$, at $95\%$ CL and with an unconstrained fit, these bounds loosen to $M_{\text{KK}}>2.5$ TeV and $M_{\text{KK}}>4.3$ TeV, respectively. Furthermore, comparing the $\text{RS}_c$ model results for all relevant Higgs decays with the LHC data indicates that $pp\to h\to Z Z^{\ast}, W W^{\ast}$ signal rates yield the most stringent bounds, such that $M_{g^{(1)}}$ less than $22.7$ $\text{TeV}\times(y_{\star}/3)$ in the brane-Higgs scenario and $13.2$ $\text{TeV}\times(y_{\star}/3)$ in the narrow bulk-Higgs case are excluded at $95\%$ probability \cite{Malm:2014gha}. Here $y_{\star}$ is an $\mathcal{O}(1)$ free parameter that is defined as the maximum allowed value for the elements of the anarchic 5D Yukawa coupling matrices such that $|(Y_f)_{ij}|\le y_{\star}$. Taking $y_{\star}=3$ value, which is implied by the perturbativity bound of the $\text{RS}_c$ model, leads to much stronger constraints on $M_{g^{(1)}}$ from Higgs physics than those emerging from the EW precision tests. Although these constraints can be loosen by considering smaller values of $y_{\star}$, one should keep in mind that lowering the bounds up to KK gauge bosons masses implied by EW precision constraints, $M_{g^{(1)}}=4.8$ TeV, will require too-small Yukawa couplings, $y_{\star}<0.3$ for the brane-Higgs scenario \cite{Malm:2014gha}, which will reinforce the RS flavor problem due to the enhanced corrections to $\epsilon_K$ parameter. Therefore, one is required to take moderate values of $y_{\star}$ by relatively increasing the KK scale, in order to avoid constraints from both flavor observables and Higgs physics. On the other hand, in the bulk-Higgs RS model, the study of Higgs decays and the signal strengths \cite{Archer:2014jca} shows that different fixed values of $y_{\star}$ can be considered from the range $y_{\star}\in [0.5,3]$ for the lightest KK masses upto allowed by EW precision data. Therefore keeping these constraints in mind, in our numerical analysis, we generate two sets of fundamental 5D Yukawa matrices with $y_{\star}=1.5$ and $3$, for both the RS models. Additionally, while exploring the parameter spaces of both the RS models, we apply the simultaneous constraints from $\Delta m_K$, $\epsilon_K$ and $\Delta m_{B_d}$ observables in $K^0-\smash{\overline K}^0$ and $B^0-\smash{\overline B}^0$ mixing, relevant to our study of the $\smash{\overline B}^0\to K^+\pi^-$ decay.

Next, similar to our previous analyses \cite{Lu:2016pfs,Nasrullah:2018vky}, we generate two sets of data points, for the $\text {RS}_c$ model, corresponding to anarchic 5D Yukawa matrices with $y_{\star}=1.5$ and $3$, with the nine quark bulk-mass parameters fitted to reproduce the correct values of the quark masses, CKM mixing angles and the Jarlskog determinant, all within their respective $2\sigma$ ranges. For details we refer the reader to \cite{Lu:2016pfs, Blanke:2008zb}. Similarly, for the bulk-Higgs RS model, following Refs.\cite{Archer:2014jca, Bauer:2009cf}, we generate two sets of anarchic 5D Yukawa matrices with $y_{\star}=1.5$ and 3, for a given value of $\beta$ and $M_{\text {KK}}$. Generally further lower values of $y_{\star}$ can be considered, but it is observed that for values of $y_{\star}<1$ it becomes increasingly difficult to fit the top-quark mass. Next, proper quark bulk-mass parameters $c_{Q_i}<1.5$ and $c_{q_i}<1.5$ are chosen which together with the 5D Yukawa matrices reproduce the correct values for the SM quark masses at the KK scale $\mu=1$ TeV. Also, in our study, we consider two different values of $\beta$, which correspond to different localization of the Higgs field along the extra dimension. $\beta=1$ correspond to the broad Higgs profile, while $\beta=10$ indicates the narrow Higgs profile, respectively.
\begin{figure}[!ht]
\begin{center}
\includegraphics[width=12cm]{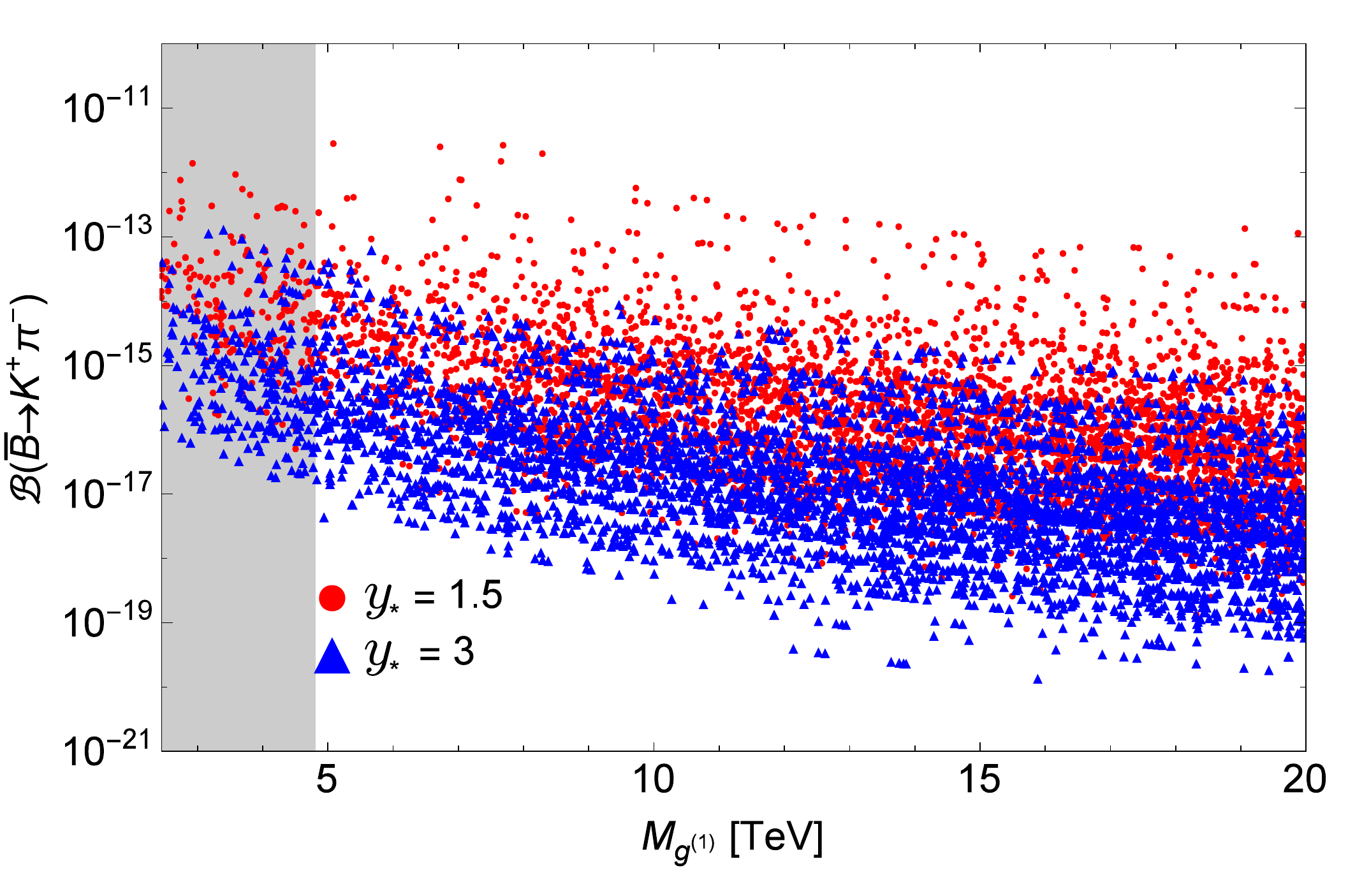}
\caption{(color online) Predictions for the $\smash{\overline B}^0\to K^+\pi^-$ branching ratio as a function of the KK gluon mass $M_{g^{(1)}}$ in the $\text{RS}_c$ model for two different values of $y_{\star}$. The gray region is excluded by the analysis of electroweak precision observables.}\label{fig:RScEW}
\end{center}
\end{figure}

After generating the data points for the two RS models, we proceed to the numerical analysis. Figure.~\ref{fig:RScEW} shows a range of the branching fraction predictions for the $\smash{\overline B}^0\to K^+\pi^-$ decay as a function of $M_{g^{(1)}}$ with two different values of $y_{\star}$, in the $\text {RS}_c$ model. The red and blue scatter points represent the cases of $y_{\star}=1.5$ and $3$, respectively. The area shaded in gray indicates the region of parameter space excluded by the tree-level analysis of EW precision measurements. Expressions of $(m_{12}^K)_{\text{KK}}$ and $(m_{12}^B)_{\text{KK}}$ relevant to $K^0-\smash{\overline K}^0$ and $B^0-\smash{\overline B}^0$ mixing constraints, calculated in the $\text {RS}_c$ model, are given in Eqs. (4.32) and (4.33) of \cite{Blanke:2008zb}, respectively. As described previously, $y_{\star}=3$ value in the $\text {RS}_c$ model with brane Higgs case, suffers strong bounds coming from Higgs physics, therefore for the considered range of the $M_{g^{(1)}}$ in Figure.~\ref{fig:RScEW}, all the scatter points with $y_{\star}=3$ value are excluded hence we will not discuss it further. The results for this case are rather presented only for a comparison with the results of the bulk-Higgs RS model. In the $y_{\star}=1.5$ case after applying the simultaneous constraints of $\Delta m_K$, $\epsilon_K$ and $\Delta m_{B_d}$, we observe in Figure.~\ref{fig:RScEW} that a large number of the scatter points in the allowed parameter space lie in the central region for each value of $M_{g^{(1)}}$ while in comparison only small number of points lie around edges. This implies that for a given value of $M_{g^{(1)}}$, more probable predictions of the $\text {RS}_c$ model are the ones lying around the central region. For example, in Figure.~\ref{fig:RScEW}, for $M_{g^{(1)}}=13$ TeV the central predictions for the branching ratio of $\smash{\overline B}^0\to K^+\pi^-$ decay in the $\text {RS}_c$ model are between $\mathcal{O}(10^{-16}-10^{-15})$, which represent three orders of magnitude enhancement than the SM prediction. On the other hand, there exist very small number of scatter points for $M_{g^{(1)}}=13$ TeV that suggest maximum possible enhancement of the branching ratio to be of $\mathcal{O}(10^{-13})$, indicating that six orders of magnitude increase compared to the SM result is possible in the $\text {RS}_c$ model.

%=====================================================================
\begin{figure}[t!]
  \centering
  \subfigure[$\beta$=1;]{
    \includegraphics[width=3.3in]{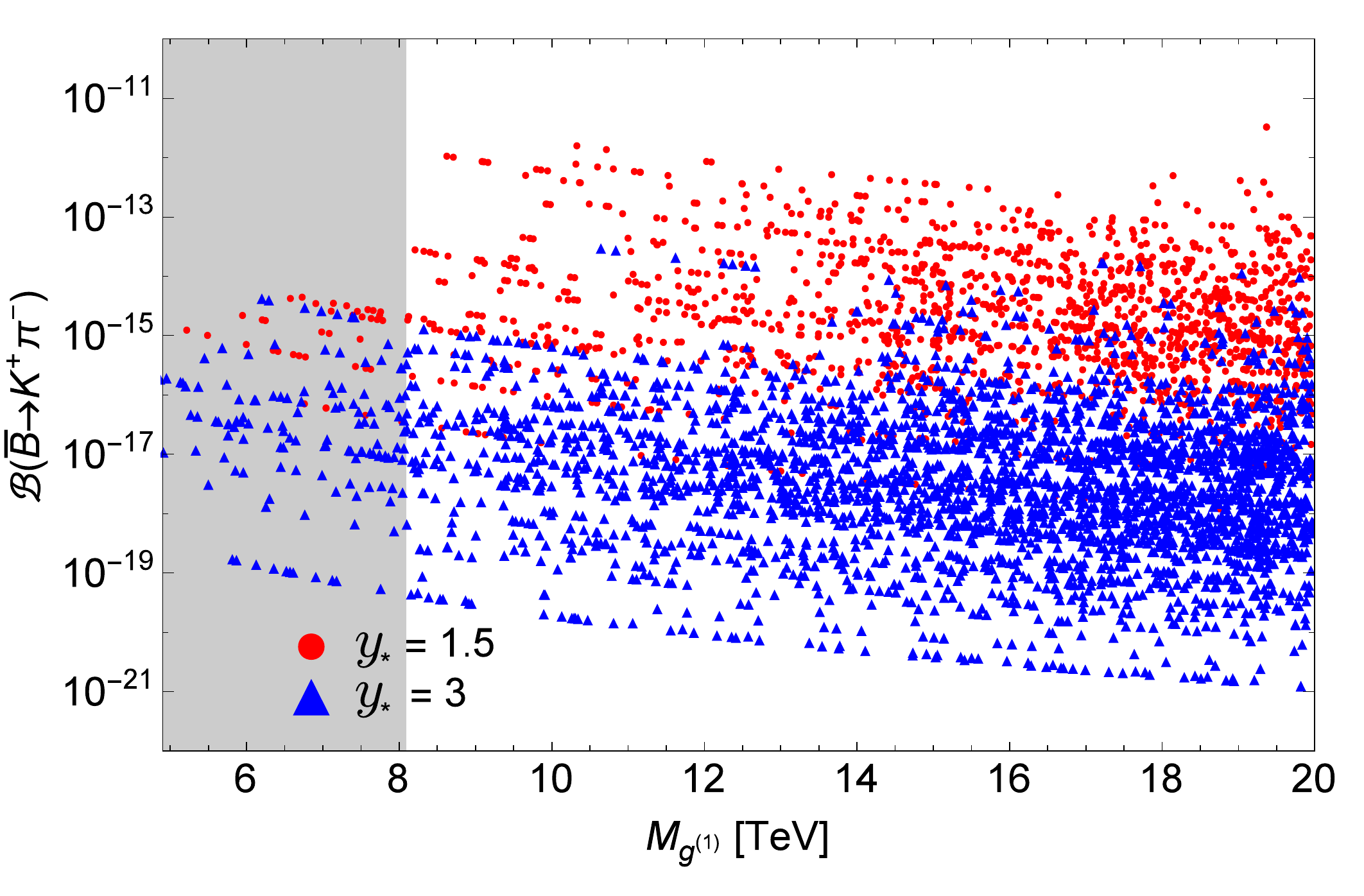}\label{fig:bHbeta1}}
  \hspace{0.in}
  \subfigure[$\beta$=10.]{
    \includegraphics[width=3.3in]{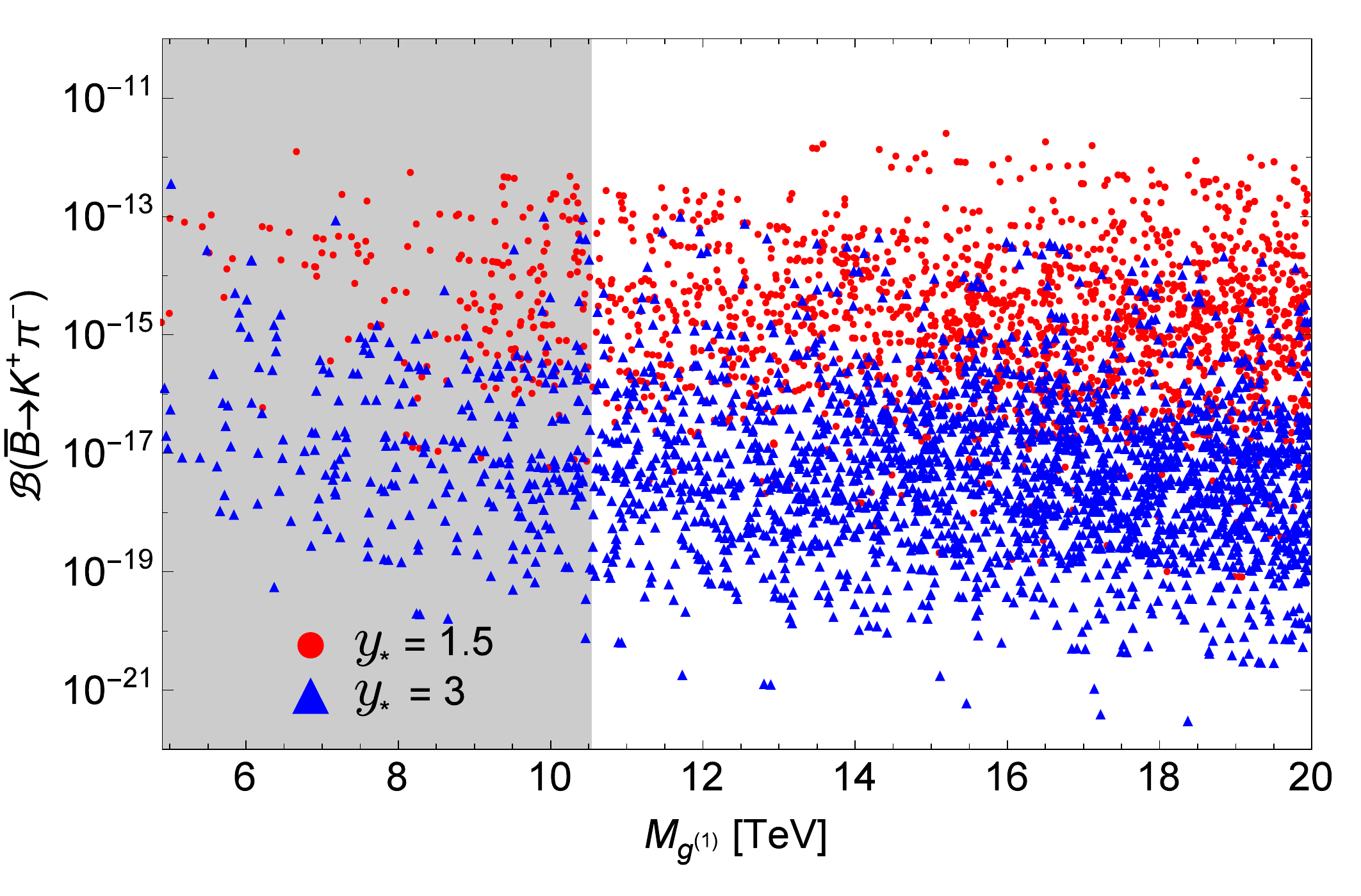}\label{fig:bHbeta10}}
  \caption{(color online) Predictions for the $\smash{\overline B}^0\to K^+\pi^-$ branching ratio as a function of the KK gluon mass $M_{g^{(1)}}$ in
  the bulk-Higgs RS model with $\beta=1$ and $\beta = 10$. The red and blue scatter points correspond to $y_{\star}=1.5$ and $3$, respectively. The gray regions are excluded by the analysis of electroweak precision observables.}
  \label{fig:bH}
\end{figure}
%======================================================================

The dominant contributions in the $\text {RS}_c$ model come from the KK gluons, whereas observing the effects of the new heavy EW gauge bosons $Z_H$ and $Z^{\prime}$, on $\smash{\overline B}^0\to K^+\pi^-$ decay, we have found in agreement with \cite{Blanke:2008zb} that while imposing the $\Delta m_K$ and $\epsilon_K$ constraints $Z_H$ and $Z^{\prime}$ give subleading contributions because the KK gluons contributions still dominate over EW contributions in $(m_{12}^K)_{\text{KK}}$ expression through the strong RG enhancement of the $C_4^{LR}$ coefficient, which only receives contribution from KK gluons, and the chiral enhancement of the $\mathcal{O}_4$ hadronic matrix element. In contrast, for the $\Delta m_{B_d}$ constraint $Z_H$ and $Z^{\prime}$ give comparable contributions to that of the first KK gluons because the RG enhancement in the $C_4^{LR}$ coefficient is smaller and the chiral enhancement of the matrix elements of $LR$ type operators is absent. Despite the fact that chiral enhancement of the matrix elements of $LR$ type operators in the $B_d$ physics observables is absent, matrix elements of the $\mathcal{O}_4$ and $\mathcal{\widetilde{O}}_{4}$ operators with $(S-P)(S+P)$ and $(S+P)(S-P)$ structures are chirally enhanced in the PQCD formalism applied for obtaining the amplitude of the $\smash{\overline B}^0\to K^+\pi^-$ process. Also as discussed previously, the WC's in the PQCD approach are calculated at the scale $t$ with $\mathcal{O}\sim (1-2)$ GeV, so the RG enhancement is also large. Both these factors play their role in increasing the $\smash{\overline B}^0\to K^+\pi^-$ branching ratio in the $\text {RS}_c$ model together with ensuring the fact that KK gluon contributions dominate over $Z_H$ and $Z^{\prime}$ contributions for a parameter point that satisfies the simultaneous constraints from $\Delta m_K$, $\epsilon_K$ and $\Delta m_{B_d}$.

In Figure.~\ref{fig:bH}, we show the predictions of the decay rate for the considered decay in the bulk-Higgs RS model for two representative values of $\beta$, after simultaneously imposing the $\Delta m_K$, $\epsilon_K$ and $\Delta m_{B_d}$ constraints. The red and blue scatter points again correspond to model points obtained using $y_{\star}=1.5$ and 3, respectively. The regions with gray shade in Figure.~\ref{fig:bH}(a) and \ref{fig:bH}(b) indicate the excluded parameter space for the $\beta=1$ and $\beta=10$ case, respectively by the analysis of electroweak precision data. Here we will mention that while imposing the experimental constraints of $\Delta m_K$, $\epsilon_K$ and $\Delta m_{B_d}$, we set the required input parameters to their central values in both the RS models and allow the resulting observables to deviate by $\pm 50\%$, $\pm 30\%$ and $\pm 30\%$, respectively in analogy to the analysis \cite{Blanke:2008zb}. Again, we see in two figures that a large number of the scatter points in the allowed parameter space lie in the central region for each value of $M_{g^{(1)}}$. $y_{\star}=3$ case with larger value correspond to more elementary fermions such that their profiles are shifted towards the UV brane, which results in more suppressed FCNC compared to the smaller value of $y_{\star}=1.5$. Therefore, from Figure.~\ref{fig:bH}, we see that the predictions of the $\smash{\overline B}^0\to K^+\pi^-$ decay rates for the parameter points with $y_{\star}=1.5$ are generally larger due to less-suppressed FCNCs than those with $y_{\star}=3$. However the smaller value of $y_{\star}$ are subjected to more severe constraints from flavor observables, hence after applying the $\Delta m_K$, $\epsilon_K$ and $\Delta m_{B_d}$ constraints simultaneously, the maximum possible $y_{\star}=1.5$ predictions reduce relatively a bit towards the case of $y_{\star}=3$. Considering maximum possible enhancement of the branching ratio in Figure.~\ref{fig:bH}(a), we note that the $y_{\star}=3$ case subject to relatively less severe constraints from the $K^0-\smash{\overline K}^0$ and $B^0-\smash{\overline B}^0$ mixing predict the branching ratio of $\mathcal{O}(10^{-14})$ for some of the parameter points while in the case of $y_{\star}=1.5$ maximum possible branching ratio is of $\mathcal{O}(10^{-13})$ for a number of parameter points, which suggest an increase of six orders of magnitude in comparison to SM prediction. Next comparing the results for $\beta=10$ case with the predictions of $\beta=1$ scenario, we observe more wider range of predictions for both cases of $y_{\star}$ in the $\beta=10$ case, such that the maximum possible branching ratios in both cases of $y_{\star}$ move further above in the $\beta=10$ scenario but in general the order of magnitude for the branching ratio remains same for both the $y_{\star}$ cases.

\section{Conclusions}\label{con}
The doubly weak $b \to dd\bar s$ transition is  highly suppressed  in the SM, which makes it sensitive to any new physics contributions beyond the SM. In this paper we have studied the pure annihilation type rare two body exclusive decay $\smash{\overline B}^0\to K^+\pi^-$, mediated by $b \to dd\bar s$ transition, within the PQCD framework. This wrong sign decay $\smash{\overline B}^0\to K^+\pi^-$ can be distinguished from the right sign decay $B^0\to K^+\pi^-$, by the time-dependent measurement of neutral B decays through $B^0-\smash{\overline B}^0$ mixing. Therefore, we propose to perform time-dependent analysis to search for the wrong sign $\smash{\overline B}^0\to K^+\pi^-$ decay, which may expose possible NP effects.

Starting with the most general local effective Hamiltonian for the $b \to dd\bar s$ processes, we analyze the $\smash{\overline B}^0\to K^+\pi^-$ decay in a model independent way, where the constraints on the Wilson coefficients of different new physics dimension-6 operators are obtained for a specific experimental precision for the observable $R$. Moreover, several examples of NP models such as NP with conserved global charge, minimal flavor violation (MFV), next-to-minimal flavor violation (NMFV) models and their predictions for the ratio $R$ are discussed. In the form of NP with a conserved charge we point out a NP mechanism, where due to the hierarchies among the NP couplings, ratio $R_X$ in different NP scenarios can be very large, after satisfying the bounds from $K^0-\smash{\overline K}^0$ and $B^0-\smash{\overline B}^0$ mixing. This mechanism represents a generalization of the sneutrino exchange in R-parity violating supersymmetry. Furthermore, sizable enhancement of the ratio $R$ is possible in NMFV models with the presence of different new physics dimension-6 operators.

We also present the results of the $\smash{\overline B}^0\to K^+\pi^-$ branching fraction in two types of RS models, after considering all the relevant constraints. In both models, the main contribution to the decay rate comes from tree level exchanges of KK gluons such that after satisfying $\Delta m_K$, $\epsilon_K$ and $\Delta m_{B_d}$ constraints simultaneously, for the $\text {RS}_c$ model with $y_{\star}=1.5$ case, a maximum increase of six orders of magnitude for the branching fraction is possible for few numbers of parameter points. Similarly, in the bulk-Higgs RS model, after considering all the relevant constraints, maximum possible enhancement of five and six orders of magnitude for the $y_{\star}=3$ and $y_{\star}=1.5$ case, respectively is probable for both broad and narrow Higgs profile
cases, which leaves this decay free for search of new physics in future experiments.

%==========================================================
\section*{Acknowledgements}\label{A5}
%==========================================================
We are grateful to Qin Qin, Wei Wang, Qi-An Zhang, Jing-Bin Liu and Chao Wang for useful discussions. The work is partly supported by National Science Foundation of China (11575151, 11521505, 11621131001) and by the Natural Science Foundation of Shandong Province (Grant No.ZR2016JL001). CDL would like to thank the financial support from the Abdus Salam International Centre for Theoretical Physics during his visit. FMB would like to acknowledge financial support from CAS-TWAS president's fellowship program 2014.

%==========================================================
%\section*{Appendix}

\begin{appendix}
%==========================================================
\section{Wave Function} \label{app-1}
The wave function describes hadronization of the quark and anti-quark to the meson $M$, which is non-perturbative but universal. For the incoming $B$ meson, its wave function is written as
 \begin{eqnarray}
\Phi_{B,\alpha\beta}(x,b) = \frac{i}{\sqrt{2N_c}}
\left[(\not \! P_1 \gamma_5)_{\alpha\beta}+ m_B \gamma_{5\alpha\beta}\right]  \phi_B(x,b),
\end{eqnarray}
where $N_c=3$ is color's degree of freedom, and $P_1$ is the its momentum. For the $\phi_B$, the model wave function has been proposed in \cite{Keum:2000ph, Keum:2000wi, Lu:2000em, Li:1992nu}
\begin{eqnarray}
\phi_B(x,b)=N_Bx^2(1-x)^2 \exp\left[-\frac{1}{2}\left(\frac{xm_B}{\omega_B}\right)^2-\frac{\omega_B^2 b^2}{2}\right] \;, \label{os}
\end{eqnarray}
where the normalization constant $N_B$ is related to the decay constant $f_B$ through
\begin{eqnarray}
\int_0^1 dx\, \phi_B(x,b=0)=\frac{f_B}{2\sqrt{2N_c}}\;.
\end{eqnarray}

The wave functions of outgoing kaon and pion, up to twist-3 accuracy, with momentum $P_2$ and $P_3$ respectively are defined as \cite{Ball:1998tj, Ball:1998je}
\begin{eqnarray}
\langle K^+(P_2)|\bar u_{\beta}(0)s_{\alpha}(z)|0\rangle=-\frac{i}{\sqrt{2N_c}}\int_0^1 \mathrm{d}x_2\, \mathrm{e}^{ix_2P_2.z} \{\gamma_5 \slashed{P}_2
\phi_K^A(x_2)+\gamma_5 m_{0K} \phi_K^P(x_2)\notag\\
-m_{0K}\gamma_5(\slashed{n}\slashed{\nu}-1) \phi_K^{T}(x_2)  \}_{\alpha\beta},
\end{eqnarray}
\begin{eqnarray}
\langle \pi^-(P_3)|\bar d_{\beta}(0)u_{\alpha}(z)|0\rangle=-\frac{i}{\sqrt{2N_c}}\int_0^1 \mathrm{d}x_3\, \mathrm{e}^{ix_3P_3.z} \{\gamma_5 \slashed{P}_3
\phi_{\pi}^A(x_3)+\gamma_5 m_{0\pi} \phi_{\pi}^P(x_3)\notag\\
-m_{0\pi}\gamma_5(\slashed{\nu}\slashed{n}-1) \phi_{\pi}^{T}(x_3)  \}_{\alpha\beta},
\end{eqnarray}
where $n=(1,0,{\bf 0}_T )\propto P_2$, $v=(0,1,{\bf 0}_T) \propto P_3$, and the chiral masses are
\begin{eqnarray}
m_{0K} = \frac{m_K^2}{m_u + m_s},\,\,\,m_{0\pi} = \frac{m_\pi^2}{m_u + m_d}.
\end{eqnarray}
The light-cone distribution amplitudes $\phi^A_{\pi(K)}$, $\phi_{\pi(K)}^P$ and $\phi_{\pi(K)}^T$ have been studied within the QCD sum rules \cite{Ball:1998tj, Ball:1998je, Ball:2004ye}, and they are expanded by Gegenbauer polynomials,
\begin{eqnarray}
\phi_{\pi(K)}^A(x) &=& \frac{f_{\pi(K)}}{2\sqrt{2N_c}} 6x(1-x)
\left[1 + a_1^{\pi(K)} C_1^{3/2}(t) +a_2^{\pi(K)}C_2^{3/2}(t) +a_4^{\pi(K)}C_4^{3/2}(t)\right] \;,
\\
\phi^P_{\pi(K)}(x) &=& \frac{f_{\pi(K)}}{2\sqrt{2N_c}}\, \bigg[ 1
+\left(30\eta_3 -\frac{5}{2}\rho_{\pi(K)}^2\right) C_2^{1/2}(t)
   -\, 3\left\{ \eta_3\omega_3 +
\frac{9}{20}\rho_{\pi(K)}^2(1+6a_2^{\pi(K)}) \right\}
C_4^{1/2}(t) \bigg]\;,
\\
\phi^T_{\pi(K)}(x) &=& \frac{f_{\pi(K)}}{2\sqrt{2N_c}}\,
(1-2x)\bigg[ 1 + 6\left(5\eta_3 -\frac{1}{2}\eta_3\omega_3 -
\frac{7}{20} \rho_{\pi(K)}^2 - \frac{3}{5}\rho_{\pi(K)}^2 a_2^{\pi(K)} \right)
(1-10x+10x^2) \bigg]\;,\ \ \ \
\end{eqnarray}
with $\rho_{\pi(K)}\equiv m_{\pi(K)}/m_{0\pi(K)}$. The Gegenbauer polynomials are defined as
\begin{eqnarray}
&\displaystyle C_2^{1/2}(t)\, =\, \frac{1}{2} \left(3\, t^2-1\right)
\;,&
C_4^{1/2}(t)\, =\, \frac{1}{8} \left(3-30\, t^2+35\, t^4\right) \;,
\nonumber\\
&\displaystyle C_1^{3/2}(t)\, =\, 3\, t \;, &
C_2^{3/2}(t)\, =\, \frac{3}{2} \left(5\, t^2-1\right) \;,
\;\;\;\;\;\;
C_4^{3/2}(t) \,=\, \frac{15}{8} \left(1-14\, t^2+21\, t^4\right) \;,
\end{eqnarray}
with $t=2x-1$. For the other parameters, such as $a_i$, $\eta_i$ and $\omega_i$, we refer the reader to \cite{Li:2005kt}.

\section{Analytic Formulae}\label{app-2}
Firstly, we present the auxiliary functions as
\begin{eqnarray}
&&G_0^2=(1-x_{2})x_{3}m^2_{B},\\
&&G_a^2=x_3 m^2_{B},\\
&&G_b^2=(1-x_2)m^2_{B},\\
&&G_c^2=(1-x_2)(x_1-x_3)m^2_{B},\\
&&G_d^2=(1-x_2(1-x_1-x_3))m^2_{B}.
\end{eqnarray}
The hard scale $t$ in the amplitudes is selected as the largest energy scale:
\begin{eqnarray}
t_{a}=\max\Big\{\sqrt{|G_0^2|},\sqrt{|G_a^2|}, 1/b_{2},1/b_{3}\Big\},\\
t_{b}=\max\Big\{\sqrt{|G_0^2|},\sqrt{|G_b^2|}, 1/b_{2},1/b_{3}\Big\},\\
t_{c}=\max\Big\{\sqrt{|G_0^2|},\sqrt{|G_c^2|}, 1/b_{1},1/b_{3}\Big\},\\
t_{d}=\max\Big\{\sqrt{|G_0^2|},\sqrt{|G_d^2|}, 1/b_{1},1/b_{3}\Big\}.
\end{eqnarray}
By inserting different operators, we calculated the amplitudes for the factorizable annihilation diagram in Fig. \ref{fig:diagrams}(a) and (b) and obtained
\begin{align}
F_{a1}&=4\pi C_F m_B^2f_B \int_0^1 dx_2 dx_3 \int_0^{\infty} b_2 db_2 b_3 db_3 \bigg[\Big\{x_3 \phi_{K}^A(x_2)\phi_{\pi}^A(x_3)\notag
\\
&+2r_{\pi}r_K \phi_{K}^P(x_2)\Big[\left(\phi_{\pi}^P(x_3)-\phi_{\pi}^T(x_3)\right)
+x_3\left(\phi_{\pi}^P(x_3)+\phi_{\pi}^T(x_3)\right)\Big] \Big\}\notag
\\
&\times E_{a}(t_a)h_a(x_2,x_3,b_2,b_3)S_t(x_3)-\Big\{(1-x_2)\phi_{K}^A(x_2)\phi_{\pi}^A(x_3)+4r_{\pi}r_K \phi_{K}^P(x_2)\phi_{\pi}^P(x_3)\notag
\\
&-2r_{\pi}r_K x_2 \phi_{\pi}^P(x_3)\left(\phi_{K}^P(x_2)-\phi_{K}^T(x_2)\right)\Big\}E_{a}(t_b)h_b(x_2,x_3,b_2,b_3)S_t(x_2)\bigg],
\end{align}
\begin{align}
F_{a2}&=4\pi C_F m_B^2f_B \int_0^1 dx_2 dx_3 \int_0^{\infty} b_2 db_2 b_3 db_3 \bigg[\Big\{2r_{K}\phi_{K}^P(x_2)\phi_{\pi}^A(x_3)\notag
\\
&+r_{\pi} x_3\phi_{K}^A(x_2)\left(\phi_{\pi}^P(x_3)-\phi_{\pi}^T(x_3)\right)\Big\}E_{a}(t_a)h_a(x_2,x_3,b_2,b_3)S_t(x_3)\notag
\\
&+\Big\{r_{K}(1-x_2)\phi_{\pi}^A(x_3)\left(\phi_{K}^P(x_2)+\phi_{K}^T(x_2)\right)\notag
\\
&+2r_{\pi}\phi_{K}^A(x_2)\phi_{\pi}^P(x_3)\Big\}
E_{a}(t_b)h_b(x_2,x_3,b_2,b_3)S_t(x_2)\bigg],
\end{align}
\begin{align}
F_{a3}=-\frac{1}{2}F_{a2}, \qquad F_{a4}=F_{a2}, \qquad F_{a5}=-\frac{1}{2}F_{a1},
\end{align}
where $C_F = 4/3$ is the group factor of $\mathrm{SU}(3)_c$ gauge group, and $r_i = m_{0i}/M_B$ ($i=K,\pi$). The threshold resummation $S_t(x)$ is parameterized as \cite{Li:2001ay}
\begin{eqnarray}
 S_t(x) = \frac{2^{1+2c}\Gamma(3/2 +c)}{\sqrt{\pi} \Gamma(1+c)}
\left[x(1-x)\right]^c,\quad c = 0.3.
\end{eqnarray}
The amplitudes for the nonfactorizable annihilation diagram in Fig. \ref{fig:diagrams}(c) and \ref{fig:diagrams}(d) are given by
\begin{align}
\mathcal{M}_{a1}&=8\pi C_F \frac{\sqrt{2N_c}}{N_c} m_B^2 \int_0^1 dx_1 dx_2 dx_3 \int_0^{\infty} b_1 db_1 b_3 db_3 \phi_B \bigg[\Big\{(1-x_2) \phi_{K}^A(x_2)\phi_{\pi}^A(x_3)\notag
\\
&+r_{\pi}r_K \Big[(1-x_2)(\phi_{K}^P(x_2)-\phi_{K}^T(x_2))(\phi_{\pi}^P(x_3)+\phi_{\pi}^T(x_3))\notag+x_3(\phi_{K}^P(x_2)+\phi_{K}^T(x_2))\notag
\\
&\times(\phi_{\pi}^P(x_3)-\phi_{\pi}^T(x_3))\Big]\Big\}
E_{a}^{\prime}(t_c)h_c(x_1,x_2,x_3,b_1,b_3)-\Big\{x_3\phi_{K}^A(x_2)\phi_{\pi}^A(x_3)\notag
\\
&+r_{\pi}r_K\Big[4\phi_{K}^P(x_2)\phi_{\pi}^P(x_3)-(1-x_3)(\phi_{K}^P(x_2)-\phi_{K}^T(x_2))(\phi_{\pi}^P(x_3)+\phi_{\pi}^T(x_3))\notag
\\
&-x_2(\phi_{K}^P(x_2)+\phi_{K}^T(x_2))(\phi_{\pi}^P(x_3)-\phi_{\pi}^T(x_3))\Big]\Big\}
E_{a}^{\prime}(t_d)h_d(x_1,x_2,x_3,b_1,b_3)\bigg].
\end{align}

\begin{align}
\mathcal{M}_{a2}&=4\pi C_F \frac{\sqrt{2N_c}}{N_c} m_B^2 \int_0^1 dx_1 dx_2 dx_3 \int_0^{\infty} b_1 db_1 b_3 db_3 \phi_B \bigg[\Big\{r_K(1-x_2)\notag\\
&\times\phi_{\pi}^A(x_3)(\phi_{K}^P(x_2)+\phi_{K}^T(x_2))+2r_{\pi}x_3\phi_{K}^A(x_2)\phi_{\pi}^P(x_3)\Big\}E_{a}^{\prime}(t_c)\notag\\
&\times h_c(x_1,x_2,x_3,b_1,b_3)+\Big\{r_K\phi_{\pi}^A(x_3)(\phi_{K}^P(x_2)+\phi_{K}^T(x_2))-2r_K x_2\phi_{\pi}^A(x_3)\phi_{K}^P(x_2)\notag
\\
&+r_{\pi}\phi_{K}^A(x_2)\Big[(\phi_{\pi}^P(x_3)+\phi_{\pi}^T(x_3))+x_3(\phi_{\pi}^P(x_3)-\phi_{\pi}^T(x_3))\Big]
\Big\}E_{a}^{\prime}(t_d)h_d(x_1,x_2,x_3,b_1,b_3)\bigg].
\end{align}

\begin{align}
\mathcal{M}_{a3}&=4\pi C_F \frac{\sqrt{2N_c}}{N_c} m_B^2 \int_0^1 dx_1 dx_2 dx_3 \int_0^{\infty} b_1 db_1 b_3 db_3 \phi_B \bigg[\Big\{-2r_K(1-x_2)\phi_{K}^P(x_2)\phi_{\pi}^A(x_3)\notag\\
&+r_{\pi}x_3\phi_{K}^A(x_2)(\phi_{\pi}^T(x_3)-\phi_{\pi}^P(x_3))\Big\}E_{a}^{\prime}(t_c)h_c(x_1,x_2,x_3,b_1,b_3)\notag
\\
&+\Big\{r_Kx_2\phi_{\pi}^A(x_3)(\phi_{K}^P(x_2)+\phi_{K}^T(x_2))-2r_K\phi_{\pi}^A(x_3)\phi_{K}^P(x_2)\notag
\\
&+r_{\pi}\phi_{K}^A(x_2)(\phi_{\pi}^P(x_3)+\phi_{\pi}^T(x_3))-2r_{\pi}x_3\phi_{K}^A(x_2)\phi_{\pi}^P(x_3)\Big\}E_{a}^{\prime}(t_d)h_d(x_1,x_2,x_3,b_1,b_3)\bigg].
\end{align}

\begin{align}
\mathcal{M}_{a4}&=-4\pi C_F \frac{\sqrt{2N_c}}{N_c} m_B^2 \int_0^1 dx_1 dx_2 dx_3 \int_0^{\infty} b_1 db_1 b_3 db_3 \phi_B \bigg[\Big\{x_3\phi_{K}^A(x_2)\phi_{\pi}^A(x_3)\notag\\
&+r_{\pi}r_K \Big[x_3(\phi_{K}^P(x_2)-\phi_{K}^T(x_2))(\phi_{\pi}^P(x_3)+\phi_{\pi}^T(x_3))\notag\\
&+(1-x_2)(\phi_{K}^P(x_2)+\phi_{K}^T(x_2))(\phi_{\pi}^P(x_3)-\phi_{\pi}^T(x_3))\Big]\Big\}E_{a}^{\prime}(t_c)\notag\\
&\times h_c(x_1,x_2,x_3,b_1,b_3)-\Big\{(1-x_2)\phi_{K}^A(x_2)\phi_{\pi}^A(x_3)-r_{\pi}r_K\Big[-4\phi_{K}^P(x_2)\phi_{\pi}^P(x_3)\notag\\
&+x_2(\phi_{K}^P(x_2)-\phi_{K}^T(x_2))(\phi_{\pi}^P(x_3)+\phi_{\pi}^T(x_3))+(1-x_3)\notag\\
&\times(\phi_{K}^P(x_2)+\phi_{K}^T(x_2))(\phi_{\pi}^P(x_3)-\phi_{\pi}^T(x_3))\Big]\Big\}E_{a}^{\prime}(t_d)h_d(x_1,x_2,x_3,b_1,b_3)\bigg].
\end{align}
\begin{align}
\mathcal{M}_{a5}&=-4\pi C_F \frac{\sqrt{2N_c}}{N_c} m_B^2 \int_0^1 dx_1 dx_2 dx_3 \int_0^{\infty} b_1 db_1 b_3 db_3 \phi_B \bigg[\Big\{-r_K(1-x_2)\notag
\\
&\times\phi_{\pi}^A(x_3)(\phi_{K}^P(x_2)-\phi_{K}^T(x_2))+r_{\pi}x_3\phi_{K}^A(x_2)
(\phi_{\pi}^P(x_3)+\phi_{\pi}^T(x_3))\Big\}\notag
\\
&\times E_{a}^{\prime}(t_c)h_c(x_1,x_2,x_3,b_1,b_3)-\Big\{
r_K(1+x_2)\phi_{\pi}^A(x_3)(\phi_{K}^P(x_2)-\phi_{K}^T(x_2))\notag
\\
&+r_{\pi}(x_3-2)\phi_{K}^A(x_2)
(\phi_{\pi}^P(x_3)+\phi_{\pi}^T(x_3))
\Big\}
E_{a}^{\prime}(t_d)h_d(x_1,x_2,x_3,b_1,b_3)\bigg].
\end{align}
In above formulas, the functions $E_{a}(t)$ and $E_{a}^{\prime}(t)$  are
\begin{eqnarray}
&&E_{a}(t)=\alpha_s(t)\exp[-S_{K}(t)-S_{\pi}(t)];\notag\\
&&E_{a}^{\prime}(t)=\alpha_s(t)\exp[-S_{B}(t)-S_{K}(t)-S_{\pi}(t)]|_{b_2=b_3}.
\end{eqnarray}
$S_B$, $S_K$, and $S_\pi$ result from resummation   of double logarithms caused by the overlap of soft collinear gluon corrections and single logarithms due to
the renormalization of ultra-violet divergence \cite{Li:1997un}, which are defined as
\begin{gather}
S_B(t) = s(x_1P_1^+,b_1) +
\frac{5}{3} \int_{1/b_1}^t \frac{d\mu'}{\mu'} \gamma_q(\mu'), \\
S_K(t)=s(x_2P_2^+,b_2)+s((1-x_2)P_2^+,b_2)+2 \int_{1/b_2}^t \frac{d\mu'}{\mu'} \gamma_q(\mu'),\\
S_\pi(t)=s(x_3P_3^-,b_3)+s((1-x_3)P_3^-,b_3)+2 \int_{1/b_3}^t \frac{d\mu'}{\mu'} \gamma_q(\mu').
\end{gather}
$s(Q,b)$, so-called Sudakov factor, is given as \cite{Li:1999kna}
\begin{eqnarray}
s(Q,b) &=& \int_{1/b}^Q \!\! \frac{d\mu'}{\mu'} \left[\left\{\frac{2}{3}(2\gamma_E-1-\log 2)+C_F \log \frac{Q}{\mu'}
 \right\} \frac{\alpha_s(\mu')}{\pi} \right. \nonumber \\
& &\left.+ \left\{ \frac{67}{9} - \frac{\pi^2}{3} - \frac{10}{27} n_f
 + \frac{2}{3} \beta_0 \log \frac{\gamma_E}{2} \right\}
 \left( \frac{\alpha_s(\mu')}{\pi} \right)^2 \log \frac{Q}{\mu'}
 \right],
 \label{eq:SudakovExpress}
\end{eqnarray}
where $\gamma_E$ is the Euler constant, and $\gamma_q = \alpha_s/\pi$ is the quark anomalous dimension. For the strong coupling constant, we use
\begin{equation}
\alpha_s(\mu)=\frac{4 \pi}{\beta_0 \log (\mu^2 / \Lambda^2)},
\label{eq:alphas}
\end{equation}
where $\beta_0 = (33-2n_f)/3$ and $n_f$ is number of active flavor. $\Lambda=250 {\rm MeV} $ is QCD scale at $n_f=4$.

The functions $h_i$ ($i=a,b,c,d$) in the decay amplitudes arise from the propagators of the virtual quark and   gluon, which are expressed by
\begin{eqnarray}
 h_a(x_2,x_3,b_2,b_3) =\left( \frac{\pi i}{2}\right)^2 H_0^{(1)}(\sqrt{G_0^2}\, b_2)
\left\{H_0^{(1)}(\sqrt{G_a^2}\, b_2)J_0(\sqrt{G_a^2}\, b_3)
\theta(b_2 - b_3) + (b_2 \leftrightarrow b_3 ) \right\},\label{eq:propagator1}
\end{eqnarray}
\begin{eqnarray}
 h_b(x_2,x_3,b_2,b_3) =\left( \frac{\pi i}{2}\right)^2 H_0^{(1)}(\sqrt{G_0^2}\, b_3)
\left\{H_0^{(1)}(\sqrt{G_b^2}\, b_2)J_0(\sqrt{G_b^2}\, b_3)
\theta(b_2 - b_3) + (b_2 \leftrightarrow b_3 ) \right\},\label{eq:propagator2}
\end{eqnarray}
\begin{multline}
 h_c(x_1,x_2,x_3,b_1,b_3) =   \biggl\{
\frac{\pi i}{2} \mathrm{H}_0^{(1)}(\sqrt{G_0^2}\, b_1)
 \mathrm{J}_0(\sqrt{G_0^2}\, b_3) \theta(b_1-b_3)+ (b_1 \leftrightarrow b_3) \biggr\}\\
 \times\left(
\begin{matrix}
 \mathrm{K}_0(\sqrt{G_c^2} b_1), & \text{for}\quad G_c^2>0 \\
 \frac{\pi i}{2} \mathrm{H}_0^{(1)}(\sqrt{|G_c^2|}\ b_1), &
 \text{for}\quad G_c^2<0
\end{matrix}\right),\label{eq:propagator3}
\end{multline}
\begin{multline}
 h_d(x_1,x_2,x_3,b_1,b_3)=\biggl\{
\frac{\pi i}{2} \mathrm{H}_0^{(1)}(\sqrt{G_0^2}\, b_1)
 \mathrm{J}_0(\sqrt{G_0^2}\, b_3) \theta(b_1-b_3)+ (b_1 \leftrightarrow b_3) \biggr\}\\
 \times\left(
\begin{matrix}
 \mathrm{K}_0(\sqrt{G_d^2} b_1), & \text{for}\quad G_d^2>0 \\
 \frac{\pi i}{2} \mathrm{H}_0^{(1)}(\sqrt{|G_d^2|}\ b_1), & \text{for}\quad G_d^2<0
\end{matrix}\right),\label{eq:propagator4}
\end{multline}
where $\mathrm{H}_0^{(1)}(z) = \mathrm{J}_0(z) + i\, \mathrm{Y}_0(z)$.
\end{appendix}
%===============================================
\bibliographystyle{bibstyle}
\bibliography{mybibfile}

\providecommand{\href}[2]{#2}\begingroup\raggedright\begin{thebibliography}{10}

\bibitem{He:1999az}
X.-G. He, C.-L. Hsueh, and J.-Q. Shi, {\it {Constraints on the phase gamma and
  new physics from $B \to K\pi$ decays}},  {\em Phys. Rev. Lett.} {\bf 84}
  (2000) 18--21, [\href{https://arxiv.org/abs/hep-ph/9905296}{{\tt
  hep-ph/9905296}}].

\bibitem{Choudhury:1998wc}
D.~Choudhury, B.~Dutta, and A.~Kundu, {\it {A Supersymmetric resolution of the
  anomaly in charmless nonleptonic B decays}},  {\em Phys. Lett.} {\bf B456}
  (1999) 185--193, [\href{https://arxiv.org/abs/hep-ph/9812209}{{\tt
  hep-ph/9812209}}].

\bibitem{Liyingreview}
Y.~Li and C.-D. Lu, {\it {Recent anomalies in B physics}},  {\em Science
  Bulletin} {\bf 63} (2018) 267--269.

\bibitem{Grossman:1999av}
Y.~Grossman, M.~Neubert, and A.~L. Kagan, {\it {Trojan penguins and isospin
  violation in hadronic B decays}},  {\em JHEP} {\bf 10} (1999) 029,
  [\href{https://arxiv.org/abs/hep-ph/9909297}{{\tt hep-ph/9909297}}].

\bibitem{Li:2005kt}
H.-n. Li, S.~Mishima, and A.~I. Sanda, {\it {Resolution to the $B \to \pi K$
  puzzle}},  {\em Phys. Rev.} {\bf D72} (2005) 114005,
  [\href{https://arxiv.org/abs/hep-ph/0508041}{{\tt hep-ph/0508041}}].

\bibitem{Barger:2004hn}
V.~Barger, C.-W. Chiang, P.~Langacker, and H.-S. Lee, {\it {Solution to the $B
  \to \pi K$ puzzle in a flavor-changing Z-prime model}},  {\em Phys. Lett.}
  {\bf B598} (2004) 218--226, [\href{https://arxiv.org/abs/hep-ph/0406126}{{\tt
  hep-ph/0406126}}].

\bibitem{Imbeault:2008ge}
M.~Imbeault, S.~Baek, and D.~London, {\it {The $B \to \pi K$ Puzzle and
  Supersymmetry}},  {\em Phys. Lett.} {\bf B663} (2008) 410--415,
  [\href{https://arxiv.org/abs/0802.1175}{{\tt arXiv:0802.1175}}].

\bibitem{PhysRevLett.81.4313}
K.~Huitu, C.-D. L\"u, P.~Singer, and D.-X. Zhang, {\it Searching for new
  physics in
  $\mathit{b}\ensuremath{\rightarrow}\mathit{ss}\overline{\mathit{d}}$ decays},
   {\em Phys. Rev. Lett.} {\bf {\bf 81}} (1998) 4313--4316,
  [\href{https://arxiv.org/abs/hep-ph/9809566}{{\tt hep-ph/9809566}}].

\bibitem{Huitu:1998pa}
K.~Huitu, C.-D. L{\"u}, P.~Singer, and D.-X. Zhang, {\it
  {$\mathit{b}\ensuremath{\rightarrow}\mathit{ss}\overline{\mathit{d}}$ decay
  in two Higgs doublet models}},  {\em Phys. Lett.} {\bf {\bf B445}} (1999)
  394--398, [\href{https://arxiv.org/abs/hep-ph/9812253}{{\tt
  hep-ph/9812253}}].

\bibitem{Wu:2003kp}
X.-H. Wu and D.-X. Zhang, {\it {Chargino contribution to the rare decay
  $\mathit{b}\ensuremath{\rightarrow}\mathit{ss}\overline{\mathit{d}}$}},  {\em
  Phys. Lett.} {\bf {\bf B587}} (2004) 95--99,
  [\href{https://arxiv.org/abs/hep-ph/0312177}{{\tt hep-ph/0312177}}].

\bibitem{Fajfer:2001ht}
S.~Fajfer and P.~Singer, {\it {Constraints on heavy $Z^{\prime}$ couplings from
  $\Delta S = 2\quad B^- \rightarrow K^- K^- \pi^+$ decay}},  {\em Phys. Rev.}
  {\bf {\bf D65}} (2002) 017301,
  [\href{https://arxiv.org/abs/hep-ph/0110233}{{\tt hep-ph/0110233}}].

\bibitem{Fajfer:2000ny}
S.~Fajfer and P.~Singer, {\it {Search for new physics in $\Delta S = 2$
  two-body (VV, PP, VP) decays of the B- meson}},  {\em Phys. Rev.} {\bf {\bf
  D62}} (2000) 117702, [\href{https://arxiv.org/abs/hep-ph/0007132}{{\tt
  hep-ph/0007132}}].

\bibitem{Fajfer:2004fx}
S.~Fajfer, J.~F. Kamenik, and P.~Singer, {\it {New-physics scenarios in $\Delta
  S = 2$ decays of the $B_c$ meson}},  {\em Phys. Rev.} {\bf {\bf D70}} (2004)
  074022, [\href{https://arxiv.org/abs/hep-ph/0407223}{{\tt hep-ph/0407223}}].

\bibitem{Fajfer:2006av}
S.~Fajfer, J.~F. Kamenik, and N.~Kosnik, {\it
  {$\mathit{b}\ensuremath{\rightarrow}\mathit{dd}\overline{\mathit{s}}$
  transition and constraints on new physics in B- decays}},  {\em Phys. Rev.}
  {\bf D74} (2006) 034027, [\href{https://arxiv.org/abs/hep-ph/0605260}{{\tt
  hep-ph/0605260}}].

\bibitem{Pirjol:2009vz}
D.~Pirjol and J.~Zupan, {\it {Predictions for
  $\mathit{b}\ensuremath{\rightarrow}\mathit{ss}\overline{\mathit{d}}$, and
  $\mathit{b}\ensuremath{\rightarrow}\mathit{dd}\overline{\mathit{s}}$ decays
  in the SM and with new physics}},  {\em JHEP} {\bf {\bf 02}} (2010) 028,
  [\href{https://arxiv.org/abs/0908.3150}{{\tt arXiv:0908.3150}}].

\bibitem{Garmash:2003er}
{\bf Belle} Collaboration, A.~Garmash et~al., {\it {Study of B meson decays to
  three body charmless hadronic final states}},  {\em Phys. Rev.} {\bf {\bf
  D69}} (2004) 012001, [\href{https://arxiv.org/abs/hep-ex/0307082}{{\tt
  hep-ex/0307082}}].

\bibitem{Aubert:2008rr}
{\bf BaBar} Collaboration, B.~Aubert et~al., {\it {Search for the highly
  suppressed decays $B^{-} \to K^{+} \pi^{-} \pi^{-}$ and $B^{-} \to K^{-}
  K^{-} \pi^{+}$}},  {\em Phys. Rev.} {\bf {\bf D78}} (2008) 091102,
  [\href{https://arxiv.org/abs/0808.0900}{{\tt arXiv:0808.0900}}].

\bibitem{LHCb:2016rul}
{\bf LHCb} Collaboration, R.~Aaij et~al., {\it {Search for the suppressed
  decays $B^{+}\rightarrow K^{+}K^{+}\pi^{-}$ and $B^{+}\rightarrow
  \pi^{+}\pi^{+}K^{-}$}},  {\em Phys. Lett.} {\bf B765} (2017) 307--316,
  [\href{https://arxiv.org/abs/1608.01478}{{\tt arXiv:1608.01478}}].

\bibitem{Aubert:2004ei}
{\bf BaBar} Collaboration, B.~Aubert et~al., {\it {Measurement of the ratio of
  decay amplitudes for $\overline{B}^0 \to J/\psi K^{*0}$ and $B^0 \to J/\psi
  K^{*0}$}},  {\em Phys. Rev. Lett.} {\bf 93} (2004) 081801,
  [\href{https://arxiv.org/abs/hep-ex/0404005}{{\tt hep-ex/0404005}}].

\bibitem{Randall:1999ee}
L.~Randall and R.~Sundrum, {\it {A Large mass hierarchy from a small extra
  dimension}},  {\em Phys. Rev. Lett.} {\bf {\bf 83}} (1999) 3370--3373,
  [\href{https://arxiv.org/abs/hep-ph/9905221}{{\tt hep-ph/9905221}}].

\bibitem{Grossman:1999ra}
Y.~Grossman and M.~Neubert, {\it {Neutrino masses and mixings in
  nonfactorizable geometry}},  {\em Phys. Lett.} {\bf {\bf B474}} (2000)
  361--371, [\href{https://arxiv.org/abs/hep-ph/9912408}{{\tt
  hep-ph/9912408}}].

\bibitem{Agashe:2006at}
K.~Agashe, R.~Contino, L.~Da~Rold, and A.~Pomarol, {\it {A Custodial symmetry
  for $Zb \bar b$}},  {\em Phys. Lett.} {\bf {\bf B641}} (2006) 62--66,
  [\href{https://arxiv.org/abs/hep-ph/0605341}{{\tt hep-ph/0605341}}].

\bibitem{Carena:2006bn}
M.~Carena, E.~Ponton, J.~Santiago, and C.~E.~M. Wagner, {\it {Light Kaluza
  Klein States in Randall-Sundrum Models with Custodial SU(2)}},  {\em Nucl.
  Phys.} {\bf {\bf B759}} (2006) 202--227,
  [\href{https://arxiv.org/abs/hep-ph/0607106}{{\tt hep-ph/0607106}}].

\bibitem{Albrecht:2009xr}
M.~E. Albrecht, M.~Blanke, A.~J. Buras, B.~Duling, and K.~Gemmler, {\it
  {Electroweak and Flavour Structure of a Warped Extra Dimension with Custodial
  Protection}},  {\em JHEP} {\bf {\bf 09}} (2009) 064,
  [\href{https://arxiv.org/abs/0903.2415}{{\tt arXiv:0903.2415}}].

\bibitem{Biancofiore:2014wpa}
P.~Biancofiore, P.~Colangelo, and F.~De~Fazio, {\it {Rare semileptonic $B\to
  K^* \ell^+ \ell^- $ decays in RS$_c$ model}},  {\em Phys. Rev.} {\bf {\bf
  D89}} (2014), no.~9 095018, [\href{https://arxiv.org/abs/1403.2944}{{\tt
  arXiv:1403.2944}}].

\bibitem{Biancofiore:2014uba}
P.~Biancofiore, P.~Colangelo, F.~De~Fazio, and E.~Scrimieri, {\it {Exclusive $b
  \to s \nu \bar \nu$ induced transitions in RS$_c$ model}},  {\em Eur. Phys.
  J.} {\bf {\bf C75}} (2015) 134, [\href{https://arxiv.org/abs/1408.5614}{{\tt
  arXiv:1408.5614}}].

\bibitem{Archer:2014jca}
P.~R. Archer, M.~Carena, A.~Carmona, and M.~Neubert, {\it {Higgs Production and
  Decay in Models of a Warped Extra Dimension with a Bulk Higgs}},  {\em JHEP}
  {\bf {\bf 01}} (2015) 060, [\href{https://arxiv.org/abs/1408.5406}{{\tt
  arXiv:1408.5406}}].

\bibitem{Keum:2000ph}
Y.-Y. Keum, H.-n. Li, and A.~I. Sanda, {\it {Fat penguins and imaginary
  penguins in perturbative QCD}},  {\em Phys. Lett.} {\bf B504} (2001) 6--14,
  [\href{https://arxiv.org/abs/hep-ph/0004004}{{\tt hep-ph/0004004}}].

\bibitem{Keum:2000wi}
Y.~Y. Keum, H.-N. Li, and A.~I. Sanda, {\it {Penguin enhancement and $B \to K
  \pi$ decays in perturbative QCD}},  {\em Phys. Rev.} {\bf D63} (2001) 054008,
  [\href{https://arxiv.org/abs/hep-ph/0004173}{{\tt hep-ph/0004173}}].

\bibitem{Lu:2000em}
C.-D. L{\"u}, K.~Ukai, and M.-Z. Yang, {\it {Branching ratio and CP violation
  of $B \to \pi \pi$ decays in perturbative QCD approach}},  {\em Phys. Rev.}
  {\bf D63} (2001) 074009, [\href{https://arxiv.org/abs/hep-ph/0004213}{{\tt
  hep-ph/0004213}}].

\bibitem{Wang:2006ria}
W.~Wang, Y.-L. Shen, Y.~Li, and C.-D. L{\"u}, {\it {Study of scalar mesons
  $f_0(980)$ and $f_0(1500)$ from $B \to f_0(980) K$ and $B \to f_0(1500) K$
  Decays}},  {\em Phys. Rev.} {\bf D74} (2006) 114010,
  [\href{https://arxiv.org/abs/hep-ph/0609082}{{\tt hep-ph/0609082}}].

\bibitem{Li:1992nu}
H.-n. Li and G.~F. Sterman, {\it {The Perturbative pion form-factor with
  Sudakov suppression}},  {\em Nucl. Phys.} {\bf B381} (1992) 129--140.

\bibitem{Li:2001ay}
H.-n. Li, {\it {Threshold resummation for exclusive B meson decays}},  {\em
  Phys. Rev.} {\bf D66} (2002) 094010,
  [\href{https://arxiv.org/abs/hep-ph/0102013}{{\tt hep-ph/0102013}}].

\bibitem{Li:1997un}
H.-n. Li and B.~Tseng, {\it {Nonfactorizable soft gluons in nonleptonic heavy
  meson decays}},  {\em Phys. Rev.} {\bf D57} (1998) 443--451,
  [\href{https://arxiv.org/abs/hep-ph/9706441}{{\tt hep-ph/9706441}}].

\bibitem{Lu:2016pfs}
C.-D. L{\"u}, F.~Munir, and Q.~Qin, {\it {$b\to ss{\bar d} $ decay in
  Randall-Sundrum models}},  {\em Chin. Phys.} {\bf C41} (2017), no.~5 053106,
  [\href{https://arxiv.org/abs/1607.07713}{{\tt arXiv:1607.07713}}].

\bibitem{Bona:2007vi}
{\bf UTfit} Collaboration, M.~Bona et~al., {\it {Model-independent constraints
  on $\Delta F=2$ operators and the scale of new physics}},  {\em JHEP} {\bf
  03} (2008) 049, [\href{https://arxiv.org/abs/0707.0636}{{\tt
  arXiv:0707.0636}}].

\bibitem{DAmbrosio:2002vsn}
G.~D'Ambrosio, G.~F. Giudice, G.~Isidori, and A.~Strumia, {\it {Minimal flavor
  violation: An Effective field theory approach}},  {\em Nucl. Phys.} {\bf
  B645} (2002) 155--187, [\href{https://arxiv.org/abs/hep-ph/0207036}{{\tt
  hep-ph/0207036}}].

\bibitem{Agashe:2005hk}
K.~Agashe, M.~Papucci, G.~Perez, and D.~Pirjol, {\it {Next to minimal flavor
  violation}},  [\href{https://arxiv.org/abs/hep-ph/0509117}{{\tt
  hep-ph/0509117}}].

\bibitem{Csaki:2008zd}
C.~Csaki, A.~Falkowski, and A.~Weiler, {\it {The Flavor of the Composite
  Pseudo-Goldstone Higgs}},  {\em JHEP} {\bf 09} (2008) 008,
  [\href{https://arxiv.org/abs/0804.1954}{{\tt arXiv:0804.1954}}].

\bibitem{Blanke:2008zb}
M.~Blanke, A.~J. Buras, B.~Duling, S.~Gori, and A.~Weiler, {\it {$\Delta F=2$
  Observables and Fine-Tuning in a Warped Extra Dimension with Custodial
  Protection}},  {\em JHEP} {\bf {\bf 03}} (2009) 001,
  [\href{https://arxiv.org/abs/0809.1073}{{\tt arXiv:0809.1073}}].

\bibitem{Bauer:2009cf}
M.~Bauer, S.~Casagrande, U.~Haisch, and M.~Neubert, {\it {Flavor Physics in the
  Randall-Sundrum Model: II. Tree-Level Weak-Interaction Processes}},  {\em
  JHEP} {\bf {\bf 09}} (2010) 017, [\href{https://arxiv.org/abs/0912.1625}{{\tt
  arXiv:0912.1625}}].

\bibitem{Duling:2009pj}
B.~Duling, {\it {A Comparative Study of Contributions to $\epsilon_K$ in the RS
  Model}},  {\em JHEP} {\bf {\bf 05}} (2010) 109,
  [\href{https://arxiv.org/abs/0912.4208}{{\tt arXiv:0912.4208}}].

\bibitem{Becirevic:2001jj}
D.~Becirevic, M.~Ciuchini, E.~Franco, V.~Gimenez, G.~Martinelli, A.~Masiero,
  M.~Papinutto, J.~Reyes, and L.~Silvestrini, {\it {$B_d - \bar{B}_d$ mixing
  and the $B_d \to J/\psi K_s$ asymmetry in general SUSY models}},  {\em Nucl.
  Phys.} {\bf {\bf B634}} (2002) 105--119,
  [\href{https://arxiv.org/abs/hep-ph/0112303}{{\tt hep-ph/0112303}}].

\bibitem{Aebischer:2017gaw}
J.~Aebischer, M.~Fael, C.~Greub, and J.~Virto, {\it {B physics Beyond the
  Standard Model at One Loop: Complete Renormalization Group Evolution below
  the Electroweak Scale}},  {\em JHEP} {\bf 09} (2017) 158,
  [\href{https://arxiv.org/abs/1704.06639}{{\tt arXiv:1704.06639}}].

\bibitem{Bauer:2008xb}
M.~Bauer, S.~Casagrande, L.~Grunder, U.~Haisch, and M.~Neubert, {\it {Little
  Randall-Sundrum models: $\epsilon_K$ strikes again}},  {\em Phys. Rev.} {\bf
  {\bf D79}} (2009) 076001, [\href{https://arxiv.org/abs/0811.3678}{{\tt
  arXiv:0811.3678}}].

\bibitem{Aad:2015fna}
{\bf ATLAS} Collaboration, G.~Aad et~al., {\it {A search for $ t\overline{t} $
  resonances using lepton-plus-jets events in proton-proton collisions at $
  \sqrt{s}=8 $ TeV with the ATLAS detector}},  {\em JHEP} {\bf 08} (2015) 148,
  [\href{https://arxiv.org/abs/1505.07018}{{\tt arXiv:1505.07018}}].

\bibitem{Sirunyan:2017uhk}
{\bf CMS} Collaboration, A.~M. Sirunyan et~al., {\it {Search for $
  t\overline{t} $ resonances in highly boosted lepton+jets and fully hadronic
  final states in proton-proton collisions at $ \sqrt{s}=13 $ TeV}},  {\em
  JHEP} {\bf 07} (2017) 001, [\href{https://arxiv.org/abs/1704.03366}{{\tt
  arXiv:1704.03366}}].

\bibitem{Malm:2013jia}
R.~Malm, M.~Neubert, K.~Novotny, and C.~Schmell, {\it {5D Perspective on Higgs
  Production at the Boundary of a Warped Extra Dimension}},  {\em JHEP} {\bf
  {\bf 01}} (2014) 173, [\href{https://arxiv.org/abs/1303.5702}{{\tt
  arXiv:1303.5702}}].

\bibitem{Dillon:2014zea}
B.~M. Dillon and S.~J. Huber, {\it {Non-Custodial Warped Extra Dimensions at
  the LHC?}},  {\em JHEP} {\bf 06} (2015) 066,
  [\href{https://arxiv.org/abs/1410.7345}{{\tt arXiv:1410.7345}}].

\bibitem{Malm:2014gha}
R.~Malm, M.~Neubert, and C.~Schmell, {\it {Higgs Couplings and Phenomenology in
  a Warped Extra Dimension}},  {\em JHEP} {\bf {\bf 02}} (2015) 008,
  [\href{https://arxiv.org/abs/1408.4456}{{\tt arXiv:1408.4456}}].

\bibitem{Nasrullah:2018vky}
A.~Nasrullah, F.~M. Bhutta, and M.~J. Aslam, {\it {The $\Lambda_b \to \Lambda\
  (\to p \pi^-)\mu^+\mu^-$ decay in the $\text{RS}_{c}$ model}},
  [\href{https://arxiv.org/abs/1805.01393}{{\tt arXiv:1805.01393}}].

\bibitem{Ball:1998tj}
P.~Ball, {\it {$B \to \pi$ and $B \to K$ transitions from QCD sum rules on the
  light cone}},  {\em JHEP} {\bf 09} (1998) 005,
  [\href{https://arxiv.org/abs/hep-ph/9802394}{{\tt hep-ph/9802394}}].

\bibitem{Ball:1998je}
P.~Ball, {\it {Theoretical update of pseudoscalar meson distribution amplitudes
  of higher twist: The Nonsinglet case}},  {\em JHEP} {\bf 01} (1999) 010,
  [\href{https://arxiv.org/abs/hep-ph/9812375}{{\tt hep-ph/9812375}}].

\bibitem{Ball:2004ye}
P.~Ball and R.~Zwicky, {\it {New results on $B \to \pi, K, \eta$ decay
  formfactors from light-cone sum rules}},  {\em Phys. Rev.} {\bf D71} (2005)
  014015, [\href{https://arxiv.org/abs/hep-ph/0406232}{{\tt hep-ph/0406232}}].

\bibitem{Li:1999kna}
H.-n. Li and B.~Melic, {\it {Determination of heavy meson wave functions from B
  decays}},  {\em Eur. Phys. J.} {\bf C11} (1999) 695--702,
  [\href{https://arxiv.org/abs/hep-ph/9902205}{{\tt hep-ph/9902205}}].

\end{thebibliography}\endgroup
\end{document}